\documentclass[10pt,floatfix,aps,twocolumn,floats,nofootinbib,pre]{revtex4-2}
\usepackage{graphics,graphicx,epsfig}
\usepackage{epsf,epstopdf,wrapfig}
\usepackage{amsmath,amssymb}
\usepackage[dvipsnames]{xcolor}
\usepackage{tikz}
\usepackage{dsfont}

\begin{document}

\preprint{APS/123-QED}

\title{Unraveling phenotypic variation in multiscale behavioral dynamics}

\author{Gautam Sridhar\textsuperscript{a,b}}
 \author{Claire Wyart\textsuperscript{a}}%
\author{Antonio Carlos Costa\textsuperscript{a,}}%
\email{Corresponding author: antonio.costa@icm-institute.org}
\affiliation{%
$^a$Sorbonne University, Paris Brain Institute (ICM), Inserm U1127, CNRS UMR 7225, 75013 Paris, France\\
$^b$Biological Physics Theory Unit, OIST Graduate University, Okinawa 904-0495, Japan 
}%

\date{\today}

\begin{abstract}

Behavioral variability reflects differences in internal and external parameters and provides a substrate for natural selection. However, uncovering its structure is challenging because behavior evolves nonlinearly across multiple timescales. Our goal is to reconstruct phenotypic variation from finite behavioral observations without prescribing the scales at which individuals should be compared. To achieve this, we introduce a multiscale dissimilarity that compares dynamics across all statistically resolvable resolutions by constructing reduced transfer operators at progressively finer scales and weighting their differences by finite-sampling uncertainty. This dissimilarity defines a geometry of dynamical phenotypes, which we reconstruct using diffusion maps. We show that the resulting phenotypic coordinates locally follow the parameter directions to which dynamics are most sensitive. We validate our approach in a stochastic model and apply it to bacterial and larval zebrafish behavior. In bacteria, we identify run speed as a major component of inter-individual variability, associated with the chemotaxis protein CheB. In zebrafish, we find that prior prey experience reshapes behavioral dynamics and phenotypic variability, with paramecia experience concentrating fish in a prey-capture-associated region of phenotypic space. Together, these results provide a framework for identifying internal and environmental variables that structure dynamical phenotypes across scales.

\end{abstract}

\maketitle

\section*{Introduction}
\begin{figure*}
    \includegraphics{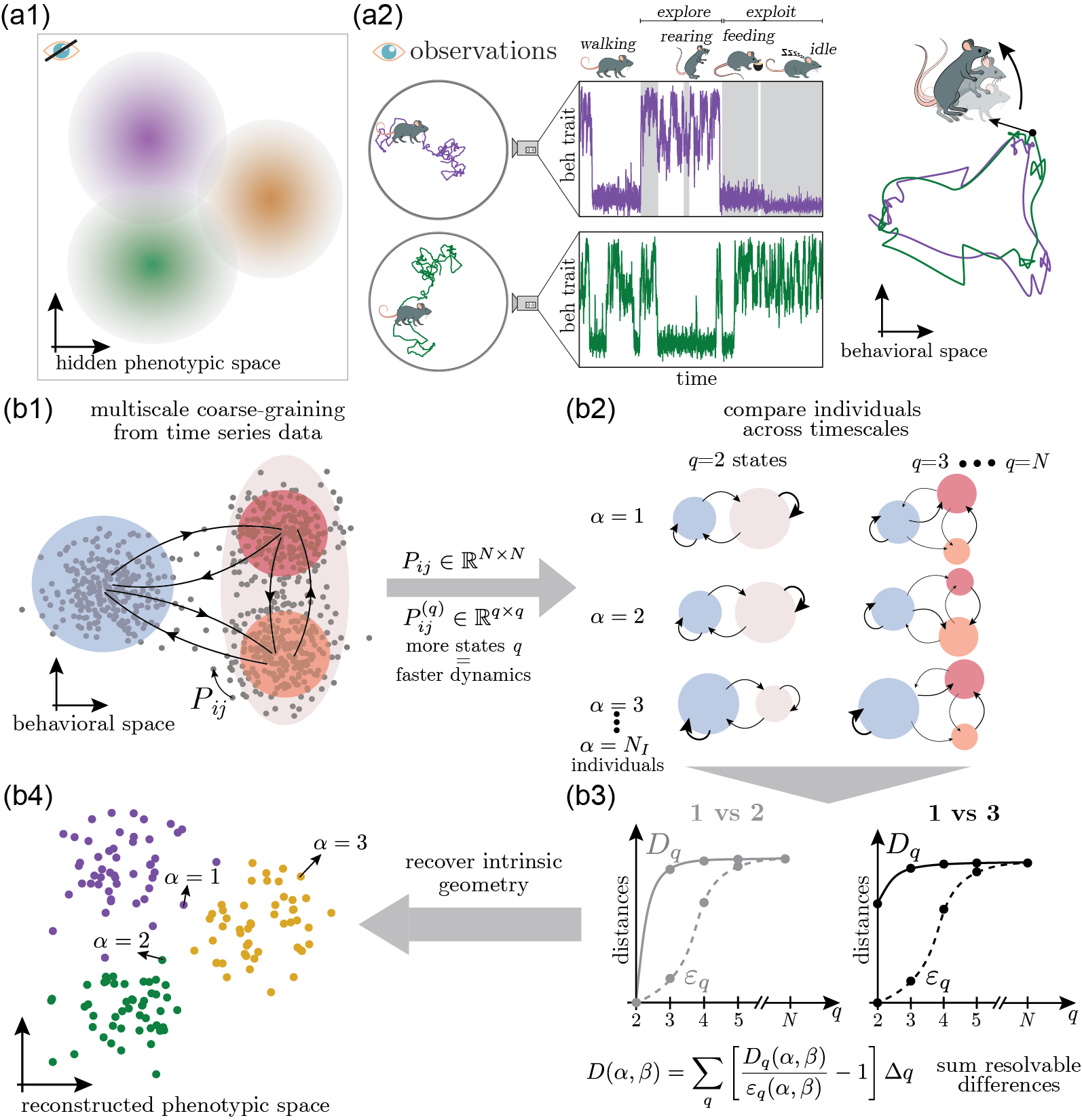}
    \caption{{\bf Unraveling behavioral variability across timescales.} 
    (a) We aim to reconstruct hidden behavioral phenotypes (a1) from behavioral observations (a2). Quantitative approaches enable the construction of rich behavioral representations  \cite{datta2019computational,berman2018measuring,brown2018ethology}. However, quantifying inter-individual differences remains a challenge as differences can manifest across timescales. For example, two individuals may use the same short time motifs, but sequence them in different ways over time. 
    (b) Our approach compares these dynamics without selecting a coarse-graining scale \emph{a priori}.
    (b1) Starting from a fine scale Markov model $P_{ij}\in\mathbb{R}^{N\times N}$ among $N$ short-time movements, we identify $q$ metastable states and construct reduced-order $q$-state models, $P_{ij}^{(q)}\in\mathbb{R}^{q\times q}$, with an increasing number of states $q$ capturing faster dynamics. 
    (b2) Each individual trajectory is projected onto the resulting family of Markov models. 
    (b3) At each resolution $q$, we measure the dynamical difference between individuals $D_q(\alpha,\beta)$, Eq.\,\ref{eq:TV_distance}, and compare it with a finite-sampling uncertainty scale $\varepsilon_q(\alpha,\beta)$, Eq.\,\ref{eq:epsilon_q}. This enables the quantification of dynamical differences across time scales. For example, individuals 1 and 2 share the same dynamics at the level of $q=2$ states, but differ in the finer scale transition at $q=3$. In contrast, individual 3 already differs from individual 1 at the level of $q=2$ states. As $q$ increases, sampling uncertainty eventually makes finer differences unresolvable. We sum the excess of $D_q(\alpha,\beta)$ over this uncertainty $\varepsilon_q(\alpha,\beta)$ across all resolvable scales to obtain a single multiscale dissimilarity, $D(\alpha,\beta)$, Eq.\,\ref{eq:D_multiscale}.
    (b4) Diffusion maps reconstruct the intrinsic geometry induced by these pairwise dissimilarities, yielding a phenotypic space that reflects significant dynamical differences across scales.
    }
    \label{fig:1}
\end{figure*}

Behavior at the scale of an entire organism is uniquely challenging to study as it operates across multiple spatial and temporal scales with substantial variability, complicating efforts to extract general principles. A common approach to cope with this variability is to average it away - in neuroscience, for example, averaging measurements to discover significant trends or effects of manipulations is widespread \cite{golowasch2002failure}. Yet, the variation across realizations may itself carry biological signal, reflecting differences in internal state \cite{pantoja2016neuromodulatory} that shape how individuals respond variably to the same input \cite{loscher2024mice}. Phenotypic variation also constitutes the substrate for natural selection: because different behavioral strategies are optimal in different ecological contexts, the spread of phenotypes in a population can shape a population’s capacity to adapt to changing environments \cite{sih2004behavioral,frankel2014adaptability}. Variability must therefore be treated not as noise to be averaged away, but as a signal reflecting the constraints that shape the parameters underlying the emergent dynamics, Fig.\,\ref{fig:1}(a1). Our goal is to reconstruct the structure of behavioral variation from finite observations, identify the dynamical scales at which individuals differ, and relate this variation to the internal and environmental variables that shape it.

While relatively static phenotypic differences can be quantified through precise measurements \cite{huttegger2011invariance}, phenotypes that are dynamic in nature, such as behavior or neural dynamics, are harder to compare. Common approaches rely on time-averaged quantities, such as mean speed or the frequency of discrete behaviors \cite{laskowski2022emergence,hernandez2021framework,deBivort2022precise}.  While insightful, these approaches miss the dynamic and multiscale nature of behavior \cite{datta2019computational,brown2018ethology,berman2018measuring}, Fig.\,\ref{fig:1}(a2). Two individuals may, for example,  have similar exploration vs. exploitation propensities, but vary in the finer scale movements used throughout. Comparing dynamical phenotypes therefore requires a representation that can resolve differences across scales without prescribing in advance which scales matter.

To achieve this, we construct a dissimilarity between finite dynamical observations that compares coarse-grained dynamics across a range of resolutions, Fig.\,\ref{fig:1}(b). At each resolution, observed differences are calibrated against the uncertainty expected from finite sampling, so that only empirically resolvable dynamical structure contributes to the comparison. The resulting dissimilarity defines a nonlinear geometry of dynamical phenotypes, which we reconstruct using diffusion maps. We further show that the resulting coordinates locally follow parameter directions to which the dynamics are most sensitive (see Supplementary Information).

We illustrate the method in a stochastic model and apply it to bacterial and larval zebrafish behavior. In bacterial motility, we find that differences in run speed are a major component of inter-individual variation and are strongly associated with variation in CheB expression. In zebrafish, we find that prior prey experience reshapes both behavioral strategies and the distribution of phenotypes in the population, with distinct effects for different prey types.

\section*{Results}

\subsection*{Reconstructing phenotypic geometry from multiscale dynamics}

To compare dynamics across timescales, we build upon recent work \cite{costa2023maximally,costa2024markovian,costa2024fluctuating,sridhar2024uncovering} and approximate the dynamics through discrete versions of linear transfer operators $P_{ij}$, which can successfully encode nonlinear dynamics \cite{bollt2013applied}.  Given a set behavioral observations, we first construct an ensemble transition matrix $P_{ij}$ across all the realizations to obtain a common behavioral space, Fig.\,\ref{fig:1}(b1) (see Methods). Its non-trivial eigenvectors identify long-lived dynamical structure and provide a basis for coarse-graining into metastable states \cite{froyland2003detecting,froyland2005statistically,prinz2011markov} (see Methods). Using increasingly many eigenvectors, we obtain a family of $q$-state Markov models $P^{(q)}_{ij}$, with larger $q$ resolving progressively faster dynamics, Fig.\,\ref{fig:1}(b1,b2).

For each coarse-graining scale $q$, we measure the difference between realization $\alpha$ and $\beta$ using the average total-variation distance among the rows of their coarse-grained transition matrices,
\begin{align}
    D_q(\alpha,\beta) =  \frac{1}{2q}\sum_i^q\sum_j^q\left| P^{(q)}_{ij}(\alpha)- P^{(q)}_{ij}(\beta)\right|,
    \label{eq:TV_distance}
\end{align}
obtaining a measure of the difference between the dynamics at a specific timescale set by $q$ (the factor of 2 bounds the distance $0\leq D_q(\alpha,\beta)\leq 1$), Fig.\,\ref{fig:1}(b3).

Crucially, we then determine which differences across multiple scales $q$ are actually resolvable from finite data. As $q$ increases, trajectories are divided among more states and fewer transitions are available to estimate each element of $P^{(q)}$. Consequently, two symbolic sequences generated by the same underlying Markov model can exhibit an apparent distance. We quantify this finite-sampling contribution by resampling under the null hypothesis that both observations were generated by the same $q$-state Markov model and defining the uncertainty scale,
\begin{align}
    \varepsilon_q(\alpha,\beta) = \left\langle \frac{1}{2q}\sum_i^q\sum_j^q\left|\tilde{P}_{ij}^{(q)}(\alpha) - \tilde{P}_{ij}^{(q)}(\beta)\right|\right\rangle,
    \label{eq:epsilon_q}
\end{align}
where $\tilde{P}_{ij}^{(q)}$ represents the resampled transition matrices and the expectation value is taken across bootstrapped resamples (see Methods).

Beyond some resolution $q$, the measured distance between two individuals is no longer distinguishable from what sampling noise alone would produce, and $D_q(\alpha,\beta)\rightarrow\varepsilon_q(\alpha,\beta)$. We therefore combine scales by accumulating, across $q$, the amount by which the measured distance exceeds this uncertainty scale,
\begin{align}
    D(\alpha,\beta) = \sum_{q=2}^{Q} \left[\frac{D_q(\alpha,\beta)}{\varepsilon_q(\alpha,\beta)}-1\right]\Delta q,
    \label{eq:D_multiscale}
\end{align}
where $Q$ is taken large enough so that $D_q/\varepsilon_q\approx1$. The resulting dissimilarity thus combines all empirically resolvable dynamical differences without selecting a preferred coarse-graining in advance.

Given these dissimilarities, we next ask how phenotypic variation is structured. Individuals differ in many underlying interacting parameters (e.g. protein expression levels, neuromodulatory states) but these need not map onto dynamical differences in a simple way \cite{marder2023individual}: the same change in a parameter can have a large effect in one region of parameter space and none in another. In addition, evolutionary trade-offs may confine phenotypes to a low-dimensional, nonlinear manifold \cite{sato2020evolutionary,kaneko2024constructing}, frequently with a simplex structure \cite{huttegger2011invariance,shoval2012evolutionary,sheftel2013geometry}. We therefore expect phenotypic variation to occupy a potentially curved, low-dimensional space whose geometry reflects how parameter variation is expressed in dynamics. To reconstruct this geometry from the pairwise dissimilarities $D(\alpha,\beta)$, we use diffusion maps \cite{coifman2006diffusion}, which recover the intrinsic geometry of a set of points by constructing a random walk that hops between nearby points, Fig.\,\ref{fig:1}(b4). We use a self-tuning kernel bandwidth, so that the walk adapts to the local density of phenotypes, and a density normalization ensuring that the recovered geometry reflects the structure of the manifold rather than how densely individuals happen to sample it (see Methods). 

This choice is not merely practical. We show that the diffusion coordinates recovered in this way locally align with the dominant modes of the metric tensor mapping parameter variation onto emergent differences in dynamics (see Supplementary Information): the directions along which the embedding separates individuals most strongly are the directions in parameter space to which the dynamics are most sensitive. The geometry we reconstruct is therefore not an arbitrary visualization of the distances, but a representation of how parameter variation is expressed in behavior. Our approach thus constructs phenotypic spaces that respect the multiscale nature of behavior, weight differences by their significance, and reflect the nontrivial relationship between underlying parameters and observed dynamics. Below we illustrate each step of our approach in a toy model.

\begin{figure*}
    \includegraphics{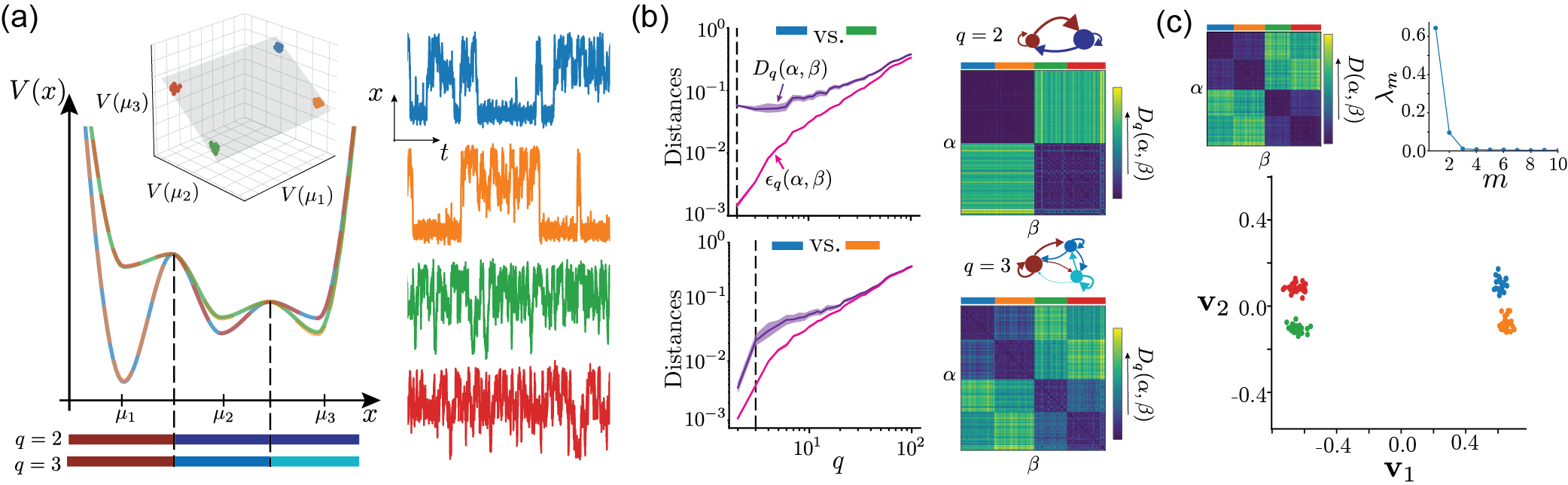}
    \caption{{\bf Reconstructing the phenotypic space of variation among overdamped stochastic dynamics in a 3-well potential.} 
    (a) We simulate overdamped Langevin dynamics in a potential with minima at $\mu_1,\mu_2,\mu_3$, and generate four dynamical ``phenotypes'' by varying the depths of the three wells. We couple $V(\mu_2)$ and $V(\mu_3)$ so that the underlying parameter variation is confined to a two-dimensional plane (inset), and show representative $200\,\text{s}$ trajectories from each phenotype (right). The colored bars below the potential indicate the coarse-grained states inferred from the dynamics: at $q=2$, the highest energy barrier separates the system into two metastable states, whereas at $q=3$ the shallower barrier is additionally resolved.
    (b) Scale-dependent dissimilarities $D_q(\alpha,\beta)$ reveal the resolution at which phenotypic differences emerge. We compare the blue phenotype with the green (top) and orange (bottom) phenotypes. Blue and green are already well separated at $q=2$, whereas distinguishing blue from orange requires the finer $q=3$ description, which resolves transitions between the two shallower wells. The corresponding coarse-grained Markov models and pairwise distance matrices for $q=2$ and $q=3$ are shown on the right. The finite-sampling uncertainty $\varepsilon_q(\alpha,\beta)$, Eq.\,\ref{eq:epsilon_q}, increases with $q$ and approaches $D_q$ for large $q$, beyond which additional differences are no longer statistically resolvable. These scale-dependent quantities are combined into the multiscale dissimilarity $D(\alpha,\beta)$, Eq.\,\ref{eq:D_multiscale}.
    (c) The resulting multiscale dissimilarity matrix (top left) retains the separation among the four phenotypes. Its diffusion-map spectrum (top right) exhibits two dominant non-trivial modes, consistent with the two-dimensional parameter variation imposed in (a). Projection onto these two diffusion coordinates (bottom) recovers the known low-dimensional organization and separates all four dynamical phenotypes.}
    \label{fig:2}
\end{figure*}

\subsubsection*{Multiscale total variation dissimilarity: a working example}

We consider overdamped dynamics in a triple-well potential, varying the depths of the three minima, $\mu_1,\mu_2,\mu_3$, to generate four dynamical ``phenotypes''. Parameter values vary weakly within each group and more strongly across groups, producing correspondingly larger differences in the emergent dynamics. We further constrain the underlying parameter variation to a two-dimensional plane, Fig.\,\ref{fig:2}(a), so that the geometry to be reconstructed is known. This model allows us to test whether the multiscale dissimilarity identifies the resolution at which dynamical differences emerge, suppresses structure that cannot be resolved from finite data, and recovers the known low-dimensional organization of the underlying parameters.

The coarse-grained models resolve the metastable structure progressively as $q$ increases, Fig.\,\ref{fig:S1_3well_cg} (see Methods). The leading non-trivial eigenvector of the reversibilized transfer operator, $\phi_1$, partitions the dynamics into $q=2$ metastable states separated by the highest energy barrier, Fig.\,\ref{fig:S1_3well_cg}(b). Including the second eigenvector, $\phi_2$, yields a $q=3$ description that additionally resolves transitions across the lower barrier, Fig.\,\ref{fig:S1_3well_cg}(c). For $q>3$, the coarse-graining further subdivides transition regions while preserving the three metastable basins, Fig.\,\ref{fig:S1_3well_cg}(d).

This scale dependence is reflected directly in the pairwise distances across $q$, Fig.\,\ref{fig:2}(b). Blue and green phenotypes are already well separated at $q=2$, whereas properly distinguishing blue from orange requires the finer $q=3$ description. Beyond $q=3$, no additional metastable structure separates the phenotypes, and finer subdivisions capturing differences in the local curvature of the potential landscape contribute little to $D_{q>3}$. At the same time, the finite-sampling uncertainty $\varepsilon_q$ increases with $q$ and approaches $D_q$ at sufficiently fine resolution. The multiscale dissimilarity, Eq.\,\ref{eq:D_multiscale}, therefore retains the resolutions at which phenotypes differ while progressively suppressing differences that become indistinguishable from sampling fluctuations. Consistently, shortening the observation time causes fine-scale differences to lose significance earlier, Fig.\,\ref{fig:S2_3well_Texpt_k}.

Finally, diffusion maps recover two dominant non-trivial coordinates, matching the two-dimensional parameter variation imposed in Fig.\,\ref{fig:2}(a). The leading coordinate $\mathbf{v}_1$ captures the larger differences associated with the left well, while $\mathbf{v}_2$ resolves the smaller differences within the right basin, separating all four phenotypic groups, Fig.\,\ref{fig:2}(c). Thus, the toy model shows that the proposed dissimilarity retains dynamical differences at the resolutions supported by the data and recovers their underlying low-dimensional organization.

\subsection*{Applications to data}

We now consider real-world datasets for which the structure of phenotypic variation may reflect variation in measurable internal mechanisms or external conditions.

\subsubsection*{From gene expression to behavioral variation in \textit{E. Coli} motility}

\begin{figure*}
    \includegraphics{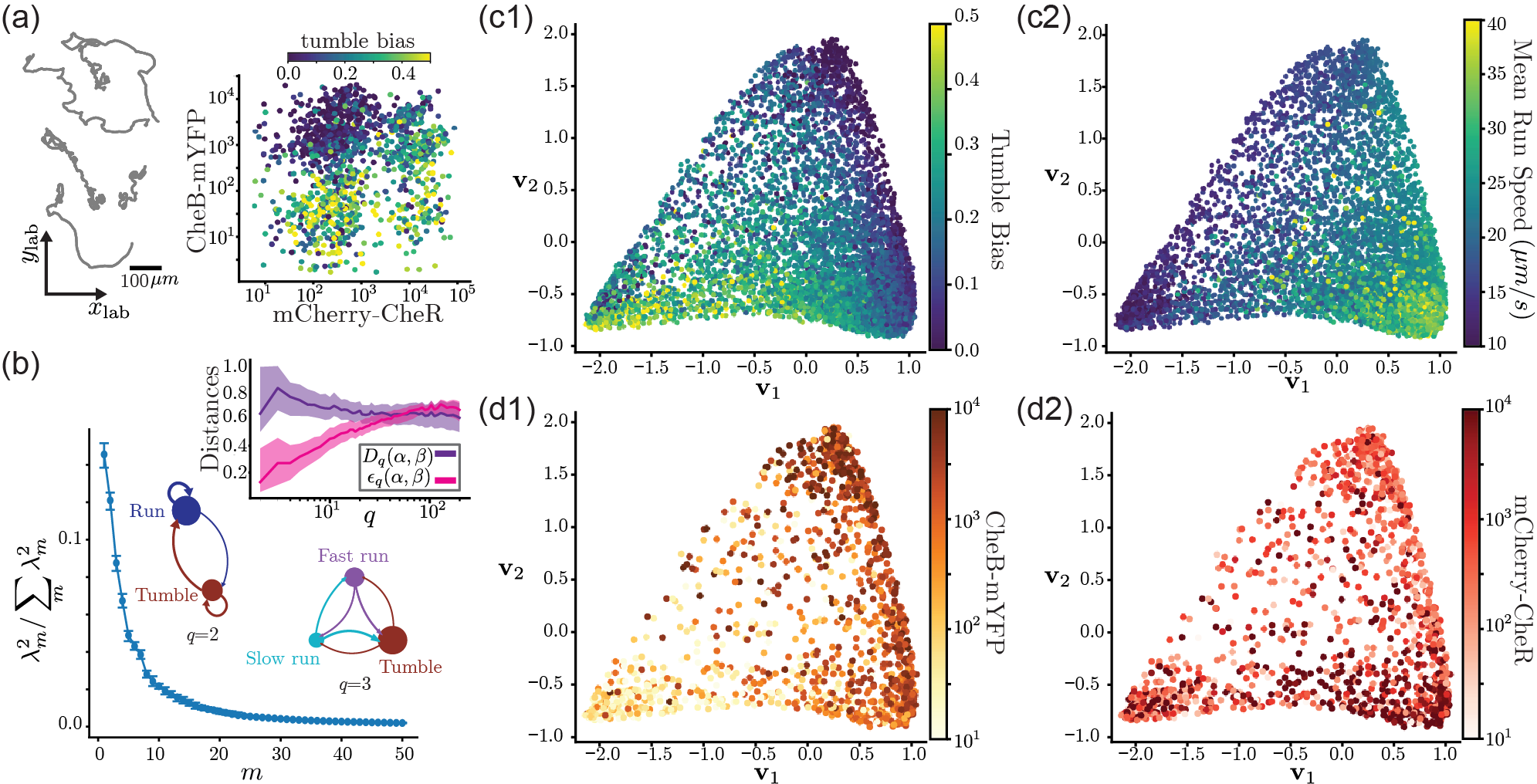}
    \caption{{\bf Phenotypic space of \textit{E. coli} motility highlights the importance of speed modulation, and the primary impact of CheB on inter-individual variability.} (a-left) Example trajectories from the dataset \cite{dufour2016direct}, which consists of RP437 isogenic wild-type bacteria ($5399$ tracks) and two mutant strains ($1374$ tracks) modified to control the expression of CheR and CheB. Trajectories last an average of $48.4\,(10.8, 191.6)\,\text{s}$. (a-right) CheR and CheB numbers for each mutant cells, color-coded by the tumble bias (as in Fig.3D of \cite{dufour2016direct}). High values of CheB and low values of CheR correspond to low tumble bias (indicative of straight runs). High values of CheR when CheB numbers are high increases the tumble bias. Low CheB values display overall higher tumble bias irrespective of variation in CheR. (b) The kinetic variance \cite{noe2015kinetic} captured by the eigenspectrum of the diffusion map reveals a structured phenotypic space where a few coordinates capture most of the kinetic variance. (b-inset)  Distances $D_q$ and uncertainties $\varepsilon_q$ between coarse-grained operators $P^{(q)}$. The differences peak at $q=3$ and are resolved by $q\approx 20$, after which the difference in phenotypes become statistically insignificant. We also show the $q=2$ and $q=3$ transition matrices for an example cell, showcasing how at $q=3$ the dynamics incorporate slower and faster runs, which are important for capturing inter-individual differences. (c) We plot the position of each cell along the two major axis of phenotypic variation, and color code them by tumble bias (c1) and mean run speed (c2). We find important difference in tumble bias along both $\mathbf{v}_1$ and $\mathbf{v}_2$, whereas variation is run speed occurs mostly along $\mathbf{v}_1$ only, see Fig.\,\ref{fig:S4_bacteria_corrs} for details. (d) Focusing on mutant strains, we find that most of the phenoypic variation aligns with differences in CheB, see also Fig.\,\ref{fig:S4_bacteria_corrs}(a). Both $\mathbf{v}_1$ and $\mathbf{v}_2$ are correlated with differences in CheB (d1), with $\mathbf{v}_2$ also showing some negative correlations with CheR when $\mathbf{v}_1$ is large (d2). This indicates that variation in CheB numbers has a primary impact on the phenotypic variability, whereas differences in CheR only becomes relevant when CheB numbers are sufficiently large.}
    \label{fig:3}
\end{figure*}

The motility of \emph{E. coli} provides a useful test case because its chemotaxis network is well characterized, while substantial phenotypic variation persists even within isogenic populations \cite{Spudich1976,Park2010} and has been linked to adaptability \cite{frankel2014adaptability,dufour2016direct}. We analyze the dataset of Dufour et al. \cite{dufour2016direct}, which combines motility trajectories from wild-type cells with measurements from strains in which the expression of the adaptation proteins CheR and CheB was controlled and quantified, Fig.\,\ref{fig:3}(a). CheR and CheB regulate receptor methylation and thereby influence CheY-P activity and tumble bias. We ask whether phenotypic variation extends beyond the canonical descriptors of mean run speed and tumble bias, and how the resulting dynamical variation relates to CheR and CheB abundance.

To reconstruct the phenotypic space, we start by inferring the long-lived dynamics from the observed center-of-mass trajectories, see Fig.\,\ref{fig:S3_bacterial_dyn} and Methods for details. We directly uncover two timescale separated behavioral modes. The slowest mode is the canonical run-and-tumble \cite{berg2004coli,Wadhams2004}, with $\approx 2s$ mixing time. The average duration of tracks is $48.4\,(10.8, 191.6)\,\text{s}$, and so the tumble bias is approximately constant \cite{plevska2021nongenetic} and constitutes the slowest timescale in the observed dynamics. The second mode differentiates fast and slow runs ($\approx 1s$ average mixing time). Beyond this, other modes have negligible timescale separation but are still interpretable: for example, the third and fourth modes capture differences in the straightness and left-right asymmetry of trajectories, Fig.\,\ref{fig:S3_bacterial_dyn}, which may however arise due to boundary effects. Overall, the $q=2$ states correspond to runs and tumble, at $q=3$ we find fast and slow runs vs tumbles, and beyond that we get finer subdivisions of these states that captures curvature and handedness differences. 

Next, we proceed to construct our multiscale dissimilarity, as described above. We find that most of the kinetic variance is captured in the first few modes of the diffusion map, Fig.\,\ref{fig:3}(b), pointing to structured variation. The multiscale dissimilarity is concentrated at low $q$, with its strongest contribution around $q=3$, and becomes indistinguishable from finite-sampling uncertainty by $q\approx 20$, Fig.\,\ref{fig:3}(b). Because the $q=3$ representation specifically resolves fast and slow runs, this peak indicates that variation in run-speed dynamics contributes substantially to differences among cells, beyond tumble bias alone, mirroring what has been found in \cite{plevska2021nongenetic}.

\begin{figure*}
    \includegraphics{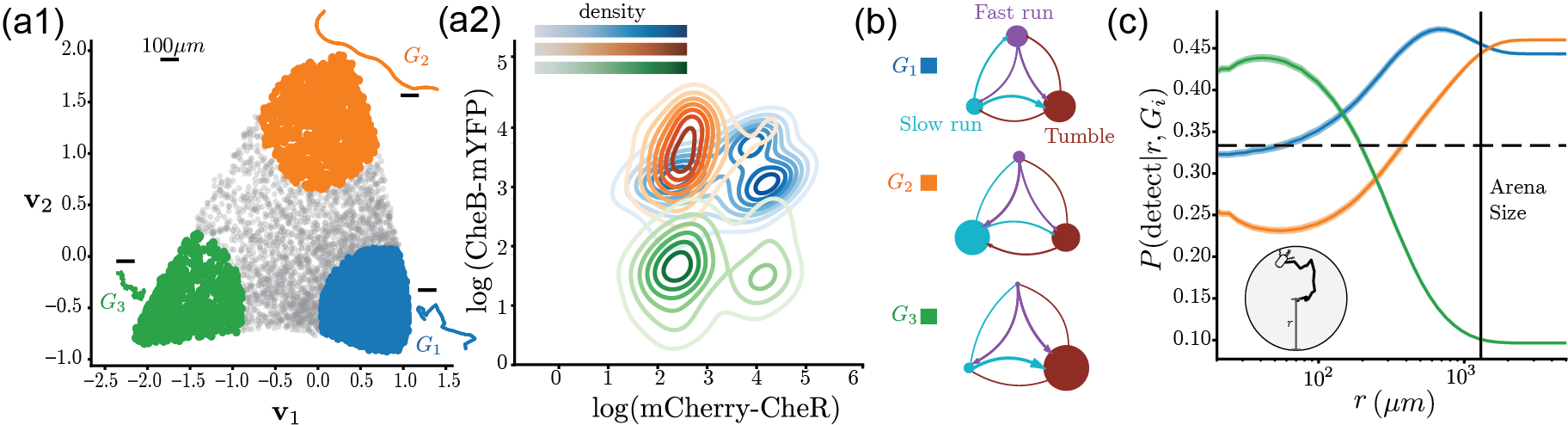}
    \caption{{\bf Speed modulation is important for the effectiveness of search strategies.} (a1) We analyze the 3 extremes of the phenotypic space in $(\mathbf{v}_1,\mathbf{v}_2)$, defined using fuzzy c-means (see Methods). We plot example trajectories from each phenotypic group. (a2) Distributions of CheR and CheB numbers for the mutant cells in each of the phenotypic groups. $G_1$ (blue) exhibits high CheB and variable CheR numbers, $G_2$ (orange) has high CheB and low CheR numbers and $G_3$ (green) has low CheB numbers and mostly low CheR numbers. (b) Average $q=3$ transition matrices in each group. Group $G_1$ ($75.9\%$ wild-type, $24.1\%$ fluorescent) showcases a bias for fast runs and tumbling, group $G_2$ ($21.4\%$ wild-type, $78.6\%$ fluorescent) exhibits more slow runs, and $G_3$ ($19.3\%$ wild-type, $80.7\%$ fluorescent) showcases high bias for tumbling. (c) The probability to detect uniformly distributed resources in a circle of radius $r$. We find that $G_3$ is most efficient at short length scales, while $G_2$ only at very large length scales (beyond the arena size of $\approx 10^3 \mu\text{m}$). These two groups also have the majority of the mutant bacteria. In contrast, $G_1$ displays close to average search efficiency at short length scales, is most efficient at length scales comparable to the arena size, and still near optimal at large length scales. This is also the group that includes the most wild-type bacteria. The $1/3$ horizontal line is the average probability over the 3 clusters.}
    \label{fig:4}
\end{figure*}

Projecting cells onto the two dominant diffusion coordinates reveals broad correspondence with canonical behavioral statistics, but not a complete reduction to them, Fig.\,\ref{fig:3}(c). A linear regression of the position along $\mathbf{v}_1$ and $\mathbf{v}_2$ from speed and tumble bias yielding an $R^2 = 0.555\,(0.546,0.567)$ (where the parenthesis represent bootstrapped 95\% confidence intervals), Fig.\,\ref{fig:S4_bacteria_corrs}(a). Nonetheless, $\mathbf{v}_1$ is strongly correlated with mean run speed, while variation along $\mathbf{v}_2$ is associated with both run speed and tumble bias. Run speeds are also more important than tumble bias in predicting the position of each individual cell in the phenotypic space, Fig.\,\ref{fig:S4_bacteria_corrs}(a1-right), indicating that run-speed modulation is a major component of inter-individual variation that is not captured by a description based on tumble bias alone.

We next asked how this phenotypic geometry relates to variation in the chemotaxis network. Among mutant cells, position in the recovered space is more strongly associated with CheB than with CheR abundance, Fig.\,\ref{fig:3}(d),Fig\.,\ref{fig:S4_bacteria_corrs}(a2). Both dominant diffusion coordinates vary systematically with CheB, whereas associations with CheR are weaker and become most apparent at high CheB abundance. This asymmetry is consistent with earlier observations in the same dataset \cite{dufour2016direct}, but the dynamical decomposition further shows that the dominant CheB-associated axis is also strongly related to run speed. Direct comparison of protein abundance with conventional behavioral statistics confirms that run speed varies systematically with CheB, in addition to the expected associations of CheR and CheB with tumble bias, Fig.\,\ref{fig:S4_bacteria_corrs}(a3).

Wild-type and mutant cells also occupy the phenotypic space differently, Fig.\,\ref{fig:S4_bacteria_corrs}(b1). The mutant population is more variable and anisotropic, with its dominant variation extending along directions associated with both tumble bias and run speed, Fig.\,\ref{fig:S4_bacteria_corrs}(b2).  Wild-type variation is more constrained and isotropic, with its major axis being strongly associated with run speed, Fig.\,\ref{fig:S4_bacteria_corrs}(b3). Thus, the directions explored by perturbing CheR and CheB do not simply reproduce the dominant variation present in the wild-type population, suggesting that additional sources of variation contribute to wild-type motility phenotypes.

We also noted that the projections along $(\mathbf{v}_1$, $\mathbf{v}_2)$ are mostly distributed inside a simplex, prompting an analysis of the differences in behavior in the extremes of the distribution, Fig.\,\ref{fig:4}(a1). We clustered the space to identify three groups (see Methods), and found that these groups differ both in protein abundance and primarily in their $q=3$ dynamics: $G_3$ has low CheB and a high tumble propensity, whereas $G_1$ (high CheB, mostly high CheR) and $G_2$ (high CheB, low CheR) are distinguished primarily by their use of fast versus slow runs, respectively, Fig.\,\ref{fig:4}(a2,b). We therefore asked whether these dynamical differences have consequences for search performance, Fig.\,\ref{fig:4}(c). Simulations of resource encounter across spatial scales show that $G_3$ performs best over short distances, whereas $G_2$ performs best only at much larger scales. $G_1$ maintains comparatively high performance across a broader range of scales and contains the largest fraction of wild-type cells, Fig.\,\ref{fig:4}(c). This suggests that the average wild-type phenotype is tuned to maintain efficient exploration at multiple length scales, in contrast with the mutant strains which become specialized to either local or global search.

\begin{figure*}
    \includegraphics{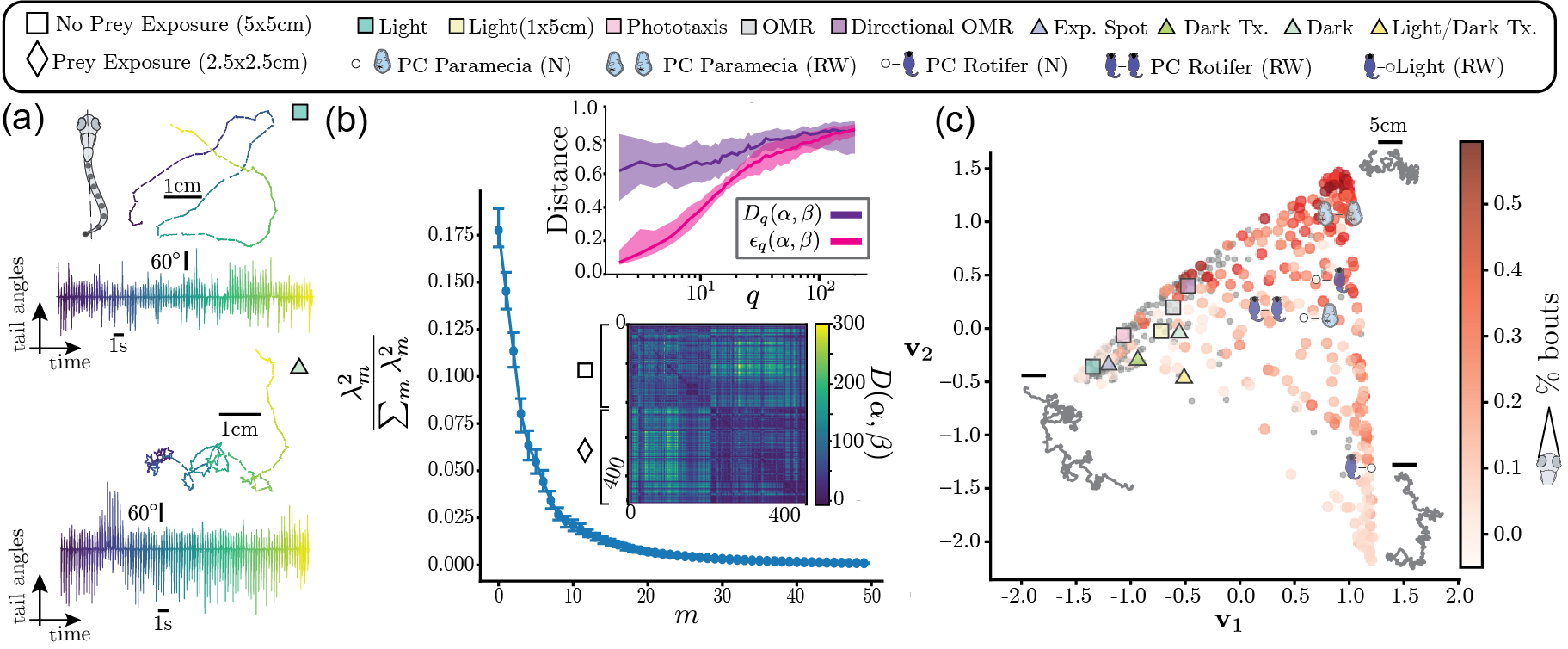}
    \caption{{\bf Phenotypic space for larval zebrafish navigation recovers prey exposure as a major determinant of variability.} We use the dataset from \cite{marques2018structure,sridhar2024uncovering}, where 463 larval zebrafish are exposed to different sensory environments (with different tick marks and colors labelled at the top).
    (a) We showcase two examples of 100 bout sequences and the corresponding fish trajectories, one from fish in the light (top) and one from fish in the dark (bottom). 
    (b) The kinetic variance captured by the eigenvalue spectrum of the diffusion map reveals a structured phenotypic space. (inset, top) Distances $D_q$ (purple) and uncertainties $\varepsilon_q$ (magenta) as a function of coarse-graining scale $q$. On average, across all sensory conditions, there is significant behavioral information up till $q\approx 20$. (inset, bottom) Multiscale dissimilarity, $D(\alpha,\beta)$, organized by fish not exposed to prey (square) and fish exposed to prey (diamond).
    (c) Projection along the first two modes of the diffusion map. Fish in hunting assays are color coded by eye convergence, a proxy for prey capture \cite{bianco2011prey}. We mark the median position in each condition with a special tick mark. The major axis of variation align with the difference between fish that were exposed to prey vs. not, as in \cite{sridhar2024uncovering}. Eye convergence is the highest in the top corner of the phenotypic space.}
    \label{fig:5}
\end{figure*}

Altogether, our approach confirmed previous findings on the importance of tumble bias and how it is controlled by CheB. In addition, we also highlighted the importance of run speeds in phenotypic variation and how it relates to CheB expression. Finally, we also found that variability in wild-type cells is not directly aligned the phenotypic axis along which CheR and CheB vary, likely reflecting the tight control of CheR and CheB numbers in wild-type cells resulting in efficient search across length scales.

\subsubsection*{Larval zebrafish navigation: multiscale variation highlights impact of prey exposure and developmental history}

To further illustrate the potential of our approach at resolving inter-individual variability, we move from bacteria to vertebrate behavior, and leverage a published dataset of larval zebrafish exposed to a variety of sensory contexts, such as aversive visual stimuli or prey of different types \cite{marques2018structure,sridhar2024uncovering}. In earlier work \cite{sridhar2024uncovering}, we used an approach similar to the one used here for uncovering a hierarchy of predictive Markov models for zebrafish behavior. However, for studying inter-individual variability we resorted to a fixed behavioral representation, which was defined by identifying a set number of coarse-grained states that could best classify all sensory contexts, which turned out to be $q=7\,\text{states}$. Here, the comparison is instead performed across all resolvable coarse-graining scales, with each resolution weighted by whether its differences exceed finite-sampling uncertainty.

In this dataset, each recording consists of a sequence of bouts, which results in distinct macroscopic movement patterns. Examining the two example behavioral sequences in Fig.\,\ref{fig:5}(a) highlights the multiscale structure of the behavior that challenges the study of inter-individual variation. Working from bout sequences, we proceed to reconstruct the phenotypic space using the multiscale dissimilarity measure introduced above. Our coarse-graining approach recovers the behavioral structure found in \cite{sridhar2024uncovering}: from cruising and wandering, to fast and slow variations, to left-right asymmetries, and so on.

We find important differences among fish up to $q\approx 20$ states, which is well beyond the $q=7$ state characterization used previously \cite{sridhar2024uncovering}. We sum all significant differences across scales, and obtain a structured phenotypic space with a few diffusion map modes capturing most of the variation, Fig.\,\ref{fig:5}(b). We project all fish along the first two dominant diffusion coordinates, Fig.\,\ref{fig:5}(c). As expected from our previous finding \cite{sridhar2024uncovering}, but this time without pre-selecting a specific coarse-graining scale, the first axis neatly separates fish never exposed to prey (negative values) and fish exposed to prey, both prior to or during the experiment (positive values). Low values of $\mathbf{v}_1$ correspond to increased cruising behavior with high speeds \cite{sridhar2024uncovering}. At high $\mathbf{v}_1$, fish exhibit two distinct behavioral patterns depending on $\mathbf{v}_2$: predominantly slow cruising and wandering with high eye convergence (a proxy for prey capture attempts) at high $\mathbf{v}_2$, or predominantly fast wandering with lower eye convergence at low $\mathbf{v}_2$.

\begin{figure*}
    \includegraphics{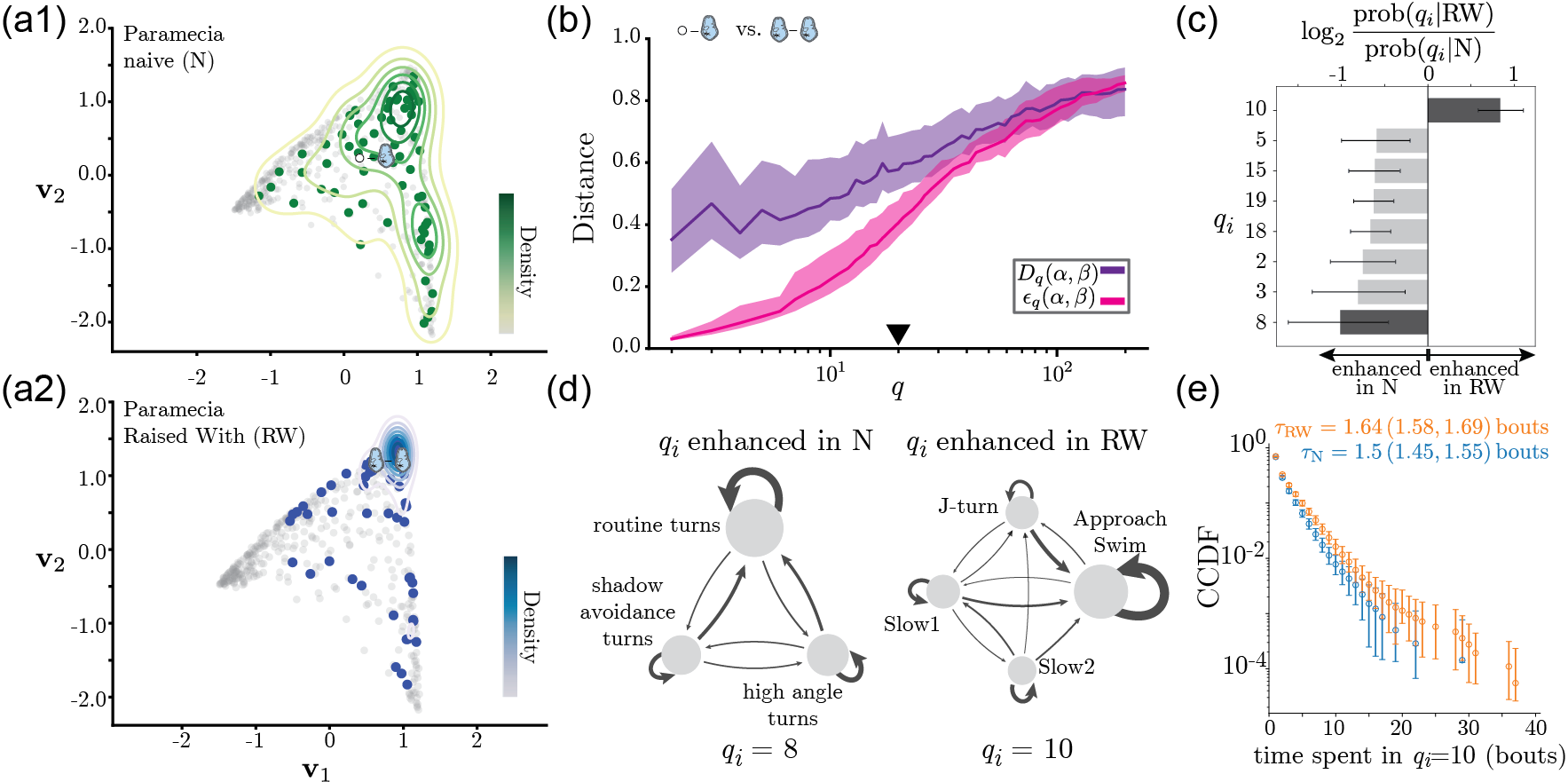}
    \caption{{\bf Prior exposure to paramecia reduces inter-individual differences and enhances prey capture sequences}
    (a1) Density plot of the fish hunting paramecia for the first time.
    (a2) Density plot of the fish hunting paramecia that were also previously raised with paramecia. Raising the fish with paramecia results in a reduction of inter-individual variability, with the distribution becoming constrained to the upper part of the phenotypic space where eye convergence is high and prey capture is most efficient, Fig.\,\ref{fig:S5_zebrafish_clus}. 
    (b) Distances $D_q$ and uncertainty scales $\varepsilon_q$ across $q$ for comparisons between naive and experience fish hunting paramecia. We find behavioral differences up till $q\approx 20$. (c) Log-fold change in the usage of each of the $q_i$ states for $q=20$. We report only states for which the distribution of bout usage was significantly differ among the two conditions (with $\text{p-value}<0.01$ in a two-sided permutation test and $|\log_2\frac{\text{prob}(q_i|\text{RW})}{\text{prob}(q_i|\text{N})}|>0.5$), and highlight the two states plotted in panel (d). (d) We plot the transition matrix among bout types \cite{marques2018structure} for the two states with the highest positive or negative log-fold change. The size of each edge represents the probability of visiting a state (we show only states with a probability $>0.1$), and arrow widths are proportional to the transition probabilities (only transitions with probability $>0.1$ are displayed). We find that $q_i=8$ is enhanced in the naive fish, consisting of sequences of routine turns, high angle turns and shadow avoidance turns \cite{marques2018structure}. In contrast, fish previously raised with paramecia exhibit an enhancement of $q_i=10$, which consists of sequences of Approach Swims and J-turns, elements of the canonical prey capture sequence \cite{mcelligott2005prey}. (e) Complementary Cumulative Distribution Function (CCDF) of the time spent in $q_i=10$ for the two conditions. We find an overall increase in bout-sequence duration following prior exposure to prey: the geometric mean duration was $\tau_\text{N} = 1.5\,(1.45,1.55)\,\text{bouts}$ for naive fish and $\tau_\text{RW} = 1.64\,(1.58,1.69)\,\text{bouts}$ for fish raised with paramecia, where values in parentheses indicate the 95\% confidence interval obtained by bootstrapping across fish. The difference between conditions was significant in a two-sided permutation test $\text{p-value}<0.0005$. The CCDF was obtained using all the available bout sequences, and error bars represent 95\% confidence intervals  bootstrapped across fish.}
    \label{fig:6}
\end{figure*}

As we found in bacteria, the first two diffusion map coordinates $\mathbf{v}_1$ and $\mathbf{v}_2$ form a simplex-like phenotypic space. We therefore examined how fish in the three extremes vary in their search and prey capture abilities (see Methods), Fig.\,\ref{fig:S5_zebrafish_clus}. Fish in the left corner are efficient at long distance exploration. Fish in the top corner efficiently search locally and capture prey directly in front of them, whereas fish in the bottom corner are effective at local search but lack frontal-capture ability. Thus, the extremes of the phenotypic space serves either the purpose of local vs global search, or efficiency at capturing prey.

To illustrate what can be learned from the full multiscale representation of zebrafish behavior, we focus on the two conditions with the largest sample sizes in the dataset: in both conditions fish are exposed the paramecia, but in the``naive'' condition fish were not previously exposed to prey ($n=65$), whereas in the ``raised with'' condition fish had previously been exposed to prey ($n=83$). Previous work has shown that prior prey experience strengthens prey-associated visual drive to the forebrain, increasing prey capture attempts and success by tuning the first turn to prey and the speed of the capture maneuver
\cite{oldfield2020experience,lagogiannis2020learning}. However, these studies did not resolve how prey exposure can alter behavior on the experimental timescale, neither how it shapes the degree of inter-individual variation among fish.

Our approach uncovers that prior exposure to paramecia reduces inter-individual differences, with a decrease in phenotypic space entropy of $H_\text{paramecia rw}-H_\text{paramecia naive} = -1\,(-1.5,-0.5)\,\text{bits, p-value}<0.0005$ (see Methods), Fig.\,\ref{fig:6}(a). Fish raised with paramecia concentrate at the region of phenotypic space where fish deploy strategies that lead to efficient frontal prey capture, Fig.\,\ref{fig:S5_zebrafish_clus}.

Focusing on the phenotypic variation across timescales, we find significant differences up till $q\approx 20$ states, Fig.\,\ref{fig:6}(b). To interpret the finest dynamical structure still supported by the data, we compare state usage in the $q=20$ representation, Fig.\,\ref{fig:6}(c),\ref{fig:S6_zebrafish_rot}(g). We find that naive fish more frequently visit states enriched in sequences of routine turns, high angle turns and shadow avoidance turns, whereas fish raised with paramecia increase their time spent in state $q_i=10$, which is enriched in sequences of Approach Swims and J-turns, the canonical prey capture sequence \cite{mcelligott2005prey,marques2018structure}, Fig.\,\ref{fig:6}(d). Not only do fish raised with prey visit state $q_i=10$ more often, but they also increase the sequence length in this state, Fig.\,\ref{fig:6}(e). Consistently, we find that fish raised with paramecia increase the time spent with eyes converged and reduce the time spent spent without eyes converged when compared to ``naive'' fish, Fig.\,\ref{fig:S6_zebrafish_rot}(e1). 

Exposure to rotifers showcases a different trend, Fig.\,\ref{fig:S6_zebrafish_rot}(c,d), with fished raised with rotifers showing an enhancement of the time spent performing exploratory behaviors (sequences of slow forward movements and routine turns), and a reduction of the visits to $q_i=10$, when compared to naive fish. They also spend more time without converged eyes, Fig.\,\ref{fig:S6_zebrafish_rot}(e2), in contrast to the pattern observed when fish were raised with paramecia. We also find that the duration of sequences in state $q_i=10$ is reduced for fish raised with rotifers, Fig.\,\ref{fig:S6_zebrafish_rot}(f). Compared to previous analysis \cite{sridhar2024uncovering}, our new approach uncovers a reduction of inter-individual differences upon prey exposure with paramecia, for which experienced fish deploy more canonical prey capture sequences.

\section*{Discussion}

This study set out to faithfully reconstruct a space of variation among dynamical phenotypes in living systems. Comparing dynamical phenotypes requires deciding which differences in time matter, yet the relevant temporal resolution is generally unknown and constrained by finite observation time. We address this problem by defining a comparison across a multiscale transfer-operator descriptions, retaining dynamical differences only to the extent that they remain distinguishable from finite-sampling uncertainty. The resulting dissimilarity therefore lets the data determine the scales at which two dynamical realizations meaningfully differ. Because this dissimilarity defines a local metric on the underlying parameter space, its global diffusion-map coordinates can be interpreted in terms of directions of dynamical sensitivity rather than merely as axes of dimensionality reduction. We illustrated the approach in a toy model and then built phenotypic spaces for bacterial and larval zebrafish navigation, confirming known features of phenotypic variation in both systems but also uncovering previously unknown effects occurring at multiple time scales.

In bacteria, we used cell-position measurements from \cite{dufour2016direct} to reconstruct a phenotypic space comprising isogenic wild-type cells together with cells in which CheR and CheB expression was experimentally varied. Phenotypic variation aligns much more strongly with CheB abundance than with CheR, consistent with the greater sensitivity to CheB reported previously \cite{dufour2016direct}. This asymmetry cannot be explained simply by the antagonistic methylation and demethylation activities of CheR and CheB: Dufour et al. showed that CheB-dependent receptor deamidation is required to reproduce the enhanced sensitivity of tumble bias to CheB abundance, while slow fluctuations in receptor methylation further contribute to CheB-dependent behavioral variability. We further find that variation in run speed constitutes a major axis distinguishing individuals, consistent with previous work identifying run speed as an important component of non-genetic behavioral diversity \cite{plevska2021nongenetic}. Unexpectedly, run speed itself is strongly associated with CheB abundance, a relationship not directly predicted by the canonical chemotaxis network and whose mechanistic origin remains unclear. Wild-type and mutant cells also occupy different regions of the phenotypic space, with distinct behavioral strategies at its extremes. Cells with experimentally varied CheR and CheB levels are enriched near phenotypes that, within our search model, perform particularly well at either short or long spatial scales. Wild-type cells are instead enriched in a region characterized by more frequent fast runs and comparatively high search performance across a broad range of scales. This raises the possibility that the native regulation of CheR and CheB helps constrain phenotypic variation away from strategies specialized to particular environmental scales, which could be advantageous when environmental length scales fluctuate unpredictably \cite{frankel2014adaptability}. Such an interpretation is consistent with the organization of the chemotaxis pathway: CheR and CheB are co-transcribed and translationally coupled, a regulatory architecture that buffers their relative expression and promotes robust chemotactic output \cite{kollmann2005design,lovdok2009role}. We emphasize, however, that the associations identified here, particularly between CheB abundance and run speed, remain correlative, and establishing their causal basis will require targeted perturbations.

In larval zebrafish, we analyzed bout-sequence data from \cite{marques2018structure}. Our previous analysis of this dataset was restricted to a single coarse-graining scale \cite{sridhar2024uncovering}. Here, by capturing coarse- and fine-scale differences within a single construction, we can resolve additional effects of prey experience on behavioral phenotypes. We found that prior exposure to paramecia substantially reduces the spread of individuals in phenotypic space. Thus, experience appears to affect not only the mean behavioral phenotype but also its variability: naive fish occupy a broader range of behavioral phenotypes, whereas prey-experienced fish converge toward a more stereotyped repertoire with bout sequences that support successful hunting. These fine-scale changes align with previous studies of hunting ontogeny: during prey capture, fish experienced to prey modulate their first turn towards the prey and the final capture maneuver to increase capture success \cite{lagogiannis2020learning,oldfield2020experience}. Our approach recovers these fine effects in bout sequences without relying on eye convergence measurements. At coarser scales, it also allows us to compare individuals over the full recording period, revealing that fish raised with paramecia spend less time performing exploratory behaviors.

Strikingly, these effects are prey-specific. In an assay with rotifers, prey-experienced fish instead spend more time performing exploratory behaviors, Fig.\,\ref{fig:S6_zebrafish_rot}, while spending less time in the prey-capture state that is enhanced after rearing with paramecia. This observation is consistent with previous reports that fish raised with rotifers spend less time hunting while nevertheless hunting more efficiently than naive fish \cite{lagogiannis2020learning}. One possible explanation is that differences in prey motility allow rotifers to be captured more rapidly, leaving more time for exploratory behavior. Differences in the nutritional value of the two prey types could also contribute to this effect. Establishing whether larvae acquire genuinely prey-specific behavioral repertoires, and identifying the mechanisms responsible for such differences, will require additional replicates and controlled manipulations of prey exposure.

Across both systems we encounter substantial structure in phenotypic variation, with implications beyond the specific organisms studied. Despite the large number of underlying parameters that can in principle vary — protein expression levels in bacteria, or membrane properties, synaptic weights and neuromodulatory states in zebrafish — the observed behavioral variation is concentrated along only a few dominant dimensions. Previous work has shown that evolutionary constraints can restrict phenotypes to low-dimensional manifolds \cite{sato2020evolutionary,kaneko2024constructing}, suggesting that population-level variation may reveal the effective directions of parameter variation that remain accessible. Furthermore, in both bacteria and zebrafish, we find that individuals are organized within simplex-like spaces whose extremes correspond to distinct behavioral strategies. Such geometries are reminiscent of Pareto-front models, in which trade-offs among competing objectives organize phenotypes between a small number of archetypal strategies \cite{shoval2012evolutionary,sheftel2013geometry}.

Methodologically, our approach builds on a rich history of reconstructing the intrinsic geometry of data points in manifold learning \cite{van2008visualizing,roweis2000nonlinear,roth2003optimal,mcinnes2018umap}, and on work dissecting the manifold structure of phenotypic spaces in evolutionary contexts \cite{huttegger2011invariance,kaneko2024constructing}. Our methodological contribution is twofold: a new multiscale dissimilarity between nonlinear dynamical systems, and a connection between the diffusion map eigenvectors and the dominant modes of the metric tensor mapping parameter variation to differences in emergent dynamics. The latter connects our work to the notion of ``sloppiness'' in model inference \cite{transtrum2015perspective}, in which parameter combinations that most strongly affect model predictions are termed ``stiff''; in the SI we show that these locally defined stiff modes guide the macroscopic modes of variation recovered by diffusion maps. 

Recent approaches compare dynamics either by learning invariant trajectory representations or by asking whether two systems are dynamically equivalent. Contrastive Cartography, for example, uses contrastive learning to construct coordinate-invariant representations of stochastic trajectories and, when applied to the same \emph{E. coli} dataset, recovered variation associated with tumble bias and CheB/CheR expression \cite{romeo2025characterizing}. Dynamical similarity methods instead quantify how far systems are from being topologically conjugate \cite{ostrow2023beyond,huang2025inputdsa,kamiya2024koopman,zhang2025koopstd}; our Perron-Frobenius operator approximation \cite{costa2023maximally} (adjoint to Koopman \cite{kutz2016dynamic}) also provides an efficient route to such comparisons, as we illustrate for the Lorenz system in the SI. Our approach addresses a complementary question: it explicitly compares coarse-grained dynamics across resolutions, calibrating each difference against finite-sampling uncertainty while retaining the identity of metastable states. This is important for behavioral variability, where individuals may visit the same states but differ in how strongly particular states are occupied \cite{werkhoven2021structure,hernandez2021framework}.

Three features of our construction bound its applicability. First, estimating transfer operators assumes that the dynamics are approximately stationary over the observation window, which is often not the case \cite{costa2024fluctuating}. We thus essentially take an adiabatic approximation of the dynamics; longer recordings may be better described as trajectories through the phenotypic space, which would in turn allow within-individual change to be studied with the same machinery.  Second, our multiscale dissimilarity requires a family of coarse-grained descriptions of the dynamics, and we obtain these with a deliberately simple and agnostic procedure based on clustering the eigenvectors of the reversibilized transfer operator and conditioning the solution on the forward coherence of each metastable state (see Methods). This was sufficient for all systems studied here, and the uncertainty scale against which we assess significant differences further regularizes against small errors in the coarse-graining. Other schemes are nonetheless available: Generalized Perron Cluster-Cluster Analysis \cite{reuter19gpcca}, for example, is constructed to handle non-reversible dynamics with cyclic structures and may be better suited to certain classes of systems. Third, the construction assumes that individuals share a common state space, so that their coarse-grained operators can be directly compared. When realizations occupy genuinely different state spaces, as in comparisons across species, dynamical similarity analysis may be more appropriate \cite{cohen2023schrodinger,germain2025spectral,kamiya2024koopman}, and our operator estimation provides an efficient route for this, as mentioned above.

Beyond these methodological limitations, the multiscale structure could also be exploited for objectives other than the unsupervised reconstruction considered here. In the present work, differences across resolutions are weighted by their statistical resolvability, without using experimental labels. When predefined conditions are available, one could instead take a supervised approach and assess which dynamical resolutions are most informative for distinguishing among conditions. 

A further caveat concerns the interpretation of the space itself. We have described the variation we recover as inter-individual, but in the datasets we studied each individual contributes a single finite-time observation, so what we strictly measure is variation among realizations. Establishing that this reflects consistent individual differences, requires repeated measurements of the same individual, the standard route in behavioral ecology \cite{laskowski2022consistent,laskowski2022emergence,werkhoven2021structure}. Our significance testing ensures that the differences we retain exceed what finite sampling of identical dynamics would produce, which rules out the trivial case, but it does not establish that an individual would return to the same position in the space if recorded again.  Nonetheless, our framework is well suited to these questions given the appropriate data. Repeated recordings would place each individual at several points in the space, allowing within-individual spread to be compared directly against between-individual spread. Following the same individuals through development would reveal whether animals occupy fixed positions, converge as they gain experience, or move along the manifold in a stereotyped way, and recording the same animals under different prey or environmental conditions would show whether the axes we recover are preserved across contexts or are specific to each assay. 

The phenotypic spaces we reconstruct are a starting point to further investigate complex mechanisms of individual variability or evolution. Because phenotypic spaces are built directly from observed dynamics, they can be aligned with independent measurements of the parameters presumed to generate variation: single-cell expression measurements in bacteria, or whole-brain imaging in larval zebrafish, where experience-dependent improvements in hunting have already been traced to identified forebrain circuits \cite{oldfield2020experience}. An individual's position in the space could then be mapped onto its internal state directly rather than inferred, and ultimately used to predict how it will respond to a perturbation rather than to describe variation after the fact. Combined with long-term evolution experiments \cite{chihoub2025evolution} and comparisons across strains and species \cite{hernandez2021framework}, this offers a route to studying evolutionarily driven behavioral variation with increased precision. In each of these settings, the variability that averaging discards is structured, and this structure is informative about the constraints shaping the parameters that control behavior.

\section*{Methods}
\subsection*{Data-driven coarse-graining of the dynamics}
To infer coarse-grained Markov models from time series data, we start by approximating the Perron-Frobenius operator using the framework of maximally predictive Markov models \cite{costa2023maximally}. When working from toy models, we directly use the full state space. When working from real data, we first reconstruct the state space using a time-delay embedding with $K$ delays. We then partition the state space into $N$ microstates $s_i, i \in [1,2,...,N]$ using a k-means clustering. We set $K$ by minimizing the unpredictability of the Markov chain (as measured through its entropy rate), and $N$ as the largest partition size before finite-size effects in the estimation of the entropy rate become evident \cite{costa2023maximally}. We then build a Markov transition matrix by counting transitions between microstates, $P_{ij}(\tau) = P(s_j(t+\tau)|s_i(t))$, yielding a discrete approximation of the Perron-Frobenius operator. We set the transition time $\tau$ such that the implied timescales $t_\text{imp}(\tau)$ of the inferred Markov chain are nearly invariant to $\tau$, signaling Markovianity \cite{costa2023maximally,costa2024markovian,sridhar2024uncovering}.

Using the fine scale transition matrices, we then proceed to identify coarse-grained metastable states. To use a common coarse-graining approach across all dynamical systems studied here, we use a simplified method that simultaneously takes advantage of the real eigenspectrum of reversibilized transition matrices, while also optimizing for the metastability of the forward dynamics. Combining ideas from  \cite{froyland2003detecting,froyland2005statistically,noe2015kinetic}, we rescale the eigenvectors of the reversibilized transition matrix by their respective eigenvalue, and then cluster this space into $q$ states using a k-means clustering weighted by the invariant measure. While we work with the real eigenspectrum of reversibilized operators, we then try to find a coarse-graining the respects the metastability of the forward dynamics by repeating the clustering from random initial conditions (1000 times), and saving the $q$ states that maximize the minimum metastability among all states. We assess metastability by estimating how much of the probability density remains in a state $q$ after a timescale $\tau$, 

\begin{equation*}\label{Eq:S_coherence}
    \chi_{\pi,\tau}(q) = \frac{\sum_{i,j\in q}\pi_i P_{ij}(\tau)}{\sum_{i \in q}  \pi_i},
\end{equation*}
where $\pi$ is the invariant density, $P\pi = \pi$.

Below we provide details of the transfer operator reconstruction for each dataset used in the main text.

{\bf Overdamped triple-well system.} To construct an ensemble transition matrix, we combine data from all parameter values. Since the dynamics is memoryless, we skip the time-delay embedding step and directly partition the measurement time series into $N=250$ for each experimental timescale $T_\text{expt}$. We then set $\tau=0.05\,\text{s}$ according to Fig.\,\ref{fig:S1_3well_cg}(a). In Fig.\,\ref{fig:S1_3well_cg}(b,c), we show the first two eigenvectors of the reversibilized transition matrix, and the obtained coarse-graining into $q=2$ and $q=3$ states that recover the wells of the system.

{\bf Bacterial motility.} To compensate for the uneven sampling of different conditions, we construct ensemble transition matrices by sampling $10^4\,\text{s}$ from both wild-type and mutant cells, and then averaging the subsequent transition matrices for each $K$ and $N$ across 100 random seeds. Using the entropy rate of the Markov chain, we find that $K^*=0.5\,\text{s}$ maximizes the predictability over the future, and $N^*=1700$ is the largest number of partitions before finite-size effects, Fig.\,\ref{fig:S3_bacterial_dyn}(a2). We choose $\tau^*=0.2\,\text{s}$ to respect the Chapman-Kolmogorov condition, Fig.\,\ref{fig:S3_bacterial_dyn}(a3).

{\bf Zebrafish behavior.} To compensate for the uneven sampling of different sensory contexts, we proceed as in \cite{sridhar2024uncovering} and to construct ensemble transition matrices by randomly sampling 7500 bouts from each condition for 100 seeds and then averaging the subsequent transition matrices. We use the same parameters as in \cite{sridhar2024uncovering}, namely $K^*=5\,\text{bouts}$ bouts, $N^*=1200$ and $\tau^*=3\,\text{bouts}$.

\subsection*{Estimating the uncertainty scale for each distance measurement}

We utilize a permutation test to determine if two realizations $\alpha,\beta$ were generated from the same underlying Markov chain at every $q$. From each pair of $q$-state symbolic sequence, $\alpha$ and $\beta$, we collect sequences of states separated by a time $\tau$. From this pool of transition events, we then randomly sample the same number of transitions as observed in each realization $\alpha$ and $\beta$. Through this procedure, we obtain resampled Markov chains, $\tilde{P}_\tau(\alpha)$ and $\tilde{P}_\tau(\beta)$, that respect the null hypothesis that the symbolic sequences in $\alpha$ and $\beta$ were generated by the same Markov model. We then estimate the average distance between resampled Markov chains using Eq.\,\ref{eq:TV_distance}, repeating this process for 100 random samples and averaging across all instances, Eq.\,\ref{eq:epsilon_q}. To avoid numerical division issues, if any $\varepsilon_q(\alpha,\beta)\approx0$, we regularize it to the maximum between the 5th percentile of the distribution of resampled distances or $10^{-5}$.

\subsection*{ Self-tuning diffusion maps}
To create a self-tuning diffusion map, we follow the methodology from \cite{zelnik2004self}. After the estimation of the multiscale dissimilarity, Eq.\,\ref{eq:D_multiscale}, $D_{\alpha,\beta}=D(\alpha,\beta)$, we transform it into a heat kernel,

\begin{equation}
    K(\alpha,\beta) = \exp\left(-\frac{D(\alpha,\beta)^2}{\sigma_\alpha\sigma_\beta}\right),
    \label{eq:diffmap_kernel}
\end{equation}
with a self-tuning bandwidth $\sigma_\alpha\sigma_\beta$ \cite{zelnik2004self}. This bandwidth is the distance to $k$-th nearest neighbor, $\sigma_{\alpha}\sigma_\beta = D_{\alpha,\alpha_k}D_{\beta,\beta_k}$, where $\alpha_k$ is $k$-th nontrivial nearest neighbor for point $\alpha$. This approach is robust to outliers while preserving local structure \cite{zelnik2004self}.

To remove the effect of the data distribution and retrieve an anisotropic kernel, we transform the above expression by dividing by the empirical estimate of the density,

\begin{align*}
    p = \sum_\alpha K(\alpha,\beta),
\end{align*}
getting the new collumn-normalized kernel,
\begin{align*}
    \hat{K}_\sigma(\alpha,\beta) = \frac{K(\alpha,\beta)}{p(\alpha)p(\beta)}.
\end{align*}
We then row-normalize this kernel to find the anisotropic transition kernel $T(\alpha,\beta)$. The two normalization steps ensure that the embedding is insensitive to how densely phenotypes are sampled, so that the recovered geometry reflects the structure of the phenotypic manifold rather than the distribution of individuals on it \cite{coifman2006diffusion}. 

We then obtain the diffusion coordinates by computing the non-trivial eigenvectors of $T(\alpha,\beta)$, $\psi_{m>0}$, normalizing them by the stationary distribution and rescaling by their respective eigenvalues $\lambda_{m>0}$, $\mathbf{v}_m = \lambda_m\psi_m$. These diffusion coordinates define the effective axes or variation and are subsequently used for visualizing and clustering the phenotypic space.

To select the appropriate $k$-th nearest neighbor we rely on inspecting the eigenvalue spectrum of $T(\alpha,\beta)$ as $k$ increases. In Fig.\,\ref{fig:S2_3well_Texpt_k}(d1-d4), we showcase how in the triple well system of Fig.\,\ref{fig:2} choosing different $k$ captures different properties of the data. A low $k=20$, the diffusion map has 3 non-trivial eigenvalues with eigenvectors that cleanly separate out the 4 clusters in the data. Such a representation is useful for spectral clustering. As $k$ increases, the global structure begins to emerge with 2 dominant non-trivial eigenvalues. In Fig.\,\ref{fig:S7_dmap_k}, we report the eigenvalue spectra for various $k$ for both the bacteria, Fig.\,\ref{fig:S7_dmap_k}(a) and larval zebrafish, Fig.\,\ref{fig:S7_dmap_k}(b), datasets. The error bars represent 95\% confidence intervals that were estimated by resampling 50\% of the data over a 100 random seeds. We set $k$ when the eigenvalue separation is sufficiently strong, which corresponds to approximately $10 \%$ of the number of data points in each dataset.

\subsection*{Discovering the phenotypic groups using fuzzy c-means}
To select out individuals in each extreme of the phenotypic space, we perform weighted fuzzy c-means clustering on the first two diffusion coordinates $(\mathbf{v}_{1},\mathbf{v}_2)$. The fuzzy c-means is weighted by density of the data points, approximated by the eigenvector of eigenvalue 1 of the diffusion operator.

\subsection*{Symbolic sequence and trajectory simulations for phenotypic groups}
We utilize the predictive power of our Markov transition matrices to simulate artificial trajectories for both the bacterial and larval zebrafish behavior. We begin by calculating ensemble Markov transition matrices for each phenotypic group $P_\tau^{(G_i)}$ utilizing only the data points belonging to group $G_i$. In order to retain all the finescale information, we estimate $P_\tau^{(G_i)}$ at the shortest $\tau$ possible for each setting. For bacterial navigation this corresponds to $\tau=0.01s$ and for larval zebrafish this corresponds to $\tau=1$ bout. We then simulate long sequences of microstates $s_i, i\in {1,...,N*}$ according to this transition matrix $P_\tau^{(G_i)}$, sampled from a random initial microstate. For bacteria, this corresponds to 10000 sequences $500\,\text{s}$ long at a sampling rate of 10Hz per phenotypic group. For larval zebrafish, this corresponds to 10000 sequences $5000\,\text{bouts}$ per phenotypic group. From each simulated sequence of microstates, we generate lab space trajectories by sampling velocity vectors from the real-data distribution of vectors within each microstate.

\subsection*{Resource detection estimation from simulated trajectories}
Using the simulated lab space trajectories, we estimate and compare the efficiency of detecting resources. For the larval zebrafish, we perform the estimation in two different conditions from \cite{sridhar2024uncovering}. In the first condition, we assume that the fish is converging its eyes and executing the prey pursuit sequence. Thus, the task is to collect resources uniformly distributed in a $60^\circ$ aperture \cite{bianco2011prey} at a maximum distance of $r=2\text{mm}$ \cite{mcelligott2005prey} from the fish. We also consider that the fish captures prey only if its head position is within a short distance of $l_{hunt}=0.05$ body lengths. In the second condition, we assume that the fish are detecting resources that are distributed densely and uniformly in a radius of $r$ around them. We assume that fish can detect resources (visually, acoustically, or mechanically) within a distance of $l_\text{detect}=1$ body length from them.

For the bacteria, we only estimate the condition of detecting resources that are distributed densely and uniformly around them in a radius $r$. While bacteria do not perform spatial sensing, we set $l_\text{detect}=10\mu m$, which roughly corresponds to the size of the bacteria plus their flagellum. We perform this calculation solely to assess the chance that the bacteria will detect changes in a fictitious chemical gradient at a distance $r$ from it.

\subsection*{Assessing variability through phenotypic space entropies}

To assess whether raising fish with prey reduces inter-individual differences, we estimated the entropy of the distribution of fish along the dominant modes ($\mathbf{v}_1$, $\mathbf{v}_2$). To do that, we approximated the probability distributions using a 2D Gaussian kernel density estimate, with a bandwidth of 0.1, which is approximately twice the median nearest neighbor distance among fish hunting paramecia. Since each condition had a different number of fish, we resampled without replacement from the condition with most fish to obtain an equal-$n$ sampling that matches the number of fish in the smaller condition. We then obtained 95\% confidence intervals by bootstrapping among equal size samples. Significance estimates were obtained using a one-sided permutation test on equal size samples. We found that for fish hunting paramecia ($n=148$ fish across the two conditions), the significance results are insensitive to the choice of bandwidth (varying the bandwidth between 0.05 and 0.3 always yields a $\text{p-value}<0.001$). However, for fish hunting rotifers ($n=46$ fish across the two conditions), the difference in entropies is not consistently significant across bandwidths, possibly reflecting the small sample size.

\subsection*{Datasets and toy models}
\subsubsection*{Overdamped Langevin dynamics from a 3 well potential}
We simulate time series from an overdamped Langevin equation of the following form, 
\begin{align*}
    dx_t = -\partial_xV(x;\mu_1,\mu_2,\mu_3)dt + \sqrt{2T_x}dW_t
\end{align*}

where $V(x;\mu_1,\mu_2,\mu_3)$ is the potential driving the dynamics, $T_x$ is the thermal noise level and $W_t$ is a Wiener process. For the phenotypes, we generate 4 unique potentials by fitting a spline using a Piecewise Cubic Hermite Interpolating Polynomial (PCHIP) to a 3 variable local minima $\theta_i = V(\mu_i)$ and 4 fixed local maxima. We set $\theta_3$ as a function of $\theta_2$, restricting variation to a plane. The height of the minima is sampled from a gaussian distribution with a standard deviation of 0.01 and varying means. For the first minimum, $V(\mu_1)$, the mean is -2 for the blue and orange phenotypes and -0.2 for the green and red phenotypes. For the second minimum, $V(\mu_2)$, the mean is -1.2 for the blue and red phenotypes and -1 for the green and orange phenotypes. The height of the third minimum is set as $V(\mu_3) = -0.98V(\mu_2) - 2.22$ for each of the phenotypes. In this way, blue and orange, and red and green share $V(\mu_1)$, whereas red and blue, and orange and green share  $V(\mu_2)$ and  $V(\mu_3)$. We then generate a variable number of realizations from each of the four phenotypes (25 realizations for the blue phenotype, 29 for the orange, 24 realizations for the green and 28 realizations for the red) with initial conditions drawn uniformly between -1 and 1 with $T_x = 0.5$, $\delta t = 0.05s$ for a total time $T_\text{expt}=10^4\text{s}$. For Fig.\,\ref{fig:S2_3well_Texpt_k}, we sample simulation durations from $T_\text{expt} = 10^{t_i}$ in the range $t_i\in [1,4]$.

\subsubsection*{Bacterial (\textit{Escherichia coli}) navigation dataset}
We utilize the dataset from \cite{dufour2016direct} where the locomotion of isogenic wild-type bacteria and mutants is recorded. In this dataset, wild-type RP437 \textit{E. coli} cells make up the wild-type dataset and are also the parental strain for all the mutants with fluorescent protein expression. The fluorescent cells are made by first genetically constructing fluorescent protein fusions CheB-mYFP and mCherry-CheR. The gene constructs are then placed under the control of two independent inducible promotors (Rhamnose or Isopropyl $\beta$-D-1-thiogalactopyranoside) on the chromosome in a strain lacking both the native cheR and cheB genes (see \cite{dufour2016direct} for details). Cells are grown to mid-exponential phase in minimal medium and then suspended in motility buffer, in which the behavior remains nearly constant for more than one hour \cite{dufour2016direct,sourjik2007vivo}. The cells are kept under a $22\,\text{mm}^2$ cover slip with $\approx 10\mu m$ in depth, confining their motion to quasi-2D. Cell motility was recorded at 10Hz temporal resolution in a 1.3mm square field of view, with a spatial resolution of 169 $\mu m$/pixel. The tracks are then extracted using custom MATLAB code \cite{dufour2016direct}. The recorded $(x,y)$ positions of the cells were used to calculate velocity autocorrelation $C_\nu$. Specifically, the velocity autocorrelation is fit using a nonlinear least squares method to the functional form $C_\nu(t) = \nu_0^2e^{(-t/\tau)\cos(\omega t)}$, where $\nu_0$ is the average cell speed, $\tau$ is the cell directional persistence and $\omega$ is the angular frequency of the circular trajectory of the cell. If $\omega$ is particularly high, it is indicative of the cells swimming close to the coverslip surface. For our analysis, we consider only tracks that are longer $10\,\text{s}$, and with a fit $\hat{\omega} < 0.05 \ \text{rad}$. For the fluorescent cells dataset, we noticed the cells appeared immobile for the last $10\,\text{s}$ of most recordings. We thus clipped the last $10\,\text{s}$ of all recordings for the fluorescent cells. We also subsequently noticed that in many cells the interaction with the cover slip was too brief to be detected by the autocorrelation criterion, but long enough for our approach to be sensitive to it. As this circling behavior does not seem indicative of naturalistic locomotion, we subsequently removed 83 tracks from the RP437 dataset and 48 tracks from the fluorescent cells dataset upon manual inspection. Finally, in total we used 5399 RP437 tracks and 1374 fluorescent cell tracks. The protein expression of the fluorescent cells was measured using single-cell fluorescence microscopy \cite{dufour2016direct}. The behavior of each cell is represented by its egocentric velocities $v_{||}$ and $v_{\perp}$ as introduced in \cite{tao2019statistical}. In brief, we use 3 points in the trajectory along time points $t_{-1},t_0,t_1$. We project the vector from $t_0$ to $t_1$ onto the vector along  $t_{-1},t_0$. Thus, $v_{||}$ represents forward of backward motion and $v_{\perp}$ represents left or right motion with respect to the motion at the previous time point. These features are then standardized across the entire dataset to have zero mean and standard deviation of 1.

\subsubsection*{Zebrafish (\textit{Danio rerio}) navigation dataset}
We leverage the dataset from \cite{marques2018structure} and previously analyzed in \cite{sridhar2024uncovering}. Wild type Tubingen larvae of ages $6-7$ days post fertilization were recorded at high temporal resolution (700Hz) at two different pixel sizes - $27\mu m$\text{/pixel} for exposed to prey conditions and $58 \mu m$\text{/pixel} for not exposed to prey conditions. 8 points on the tail of the fish are tracked during the experiment using a custom online tracking software \cite{marques2018structure}. Same as \cite{sridhar2024uncovering}, we utilize 463 fish from 14 different sensory contexts, retaining only those detected bouts that generated a velocity of at least $4 \text{mm/s}$. Each bout is a $250\,\text{ms} \times 8$ dimensional object, which we project onto 20 dimensions using principle component analysis as in \cite{sridhar2024uncovering}. We then work with the sequences of bouts to reconstruct the state space. We also utilize the eye convergence index (bout with or without eye convergence) for each bout recorded for the fish exposed to prey as a marker of hunting attempts, which is described in detail in \cite{sridhar2024uncovering,marques2018structure}. The details of each of the sensory context is provided below, from \cite{sridhar2024uncovering}:

{\bf Light }$\mathbf{5\times 5\,\text{cm}^2}$: Fish were presented with a uniform light from below with an illuminance of $1000\,\text{lm/m}^2$ in a $5\times 5\,\text{cm}^2$ squared arena with $3\,\text{mm}$ of depth. 

{\bf Light }$\mathbf{1\times 5\,\text{cm}^2}$: Fish were presented with a uniform light from below with an illuminance of $1000\,\text{lm/m}^2$ in a $1\times 5\,\text{cm}^2$ squared arena with $8\,\text{mm}$ of depth. 

{\bf Dark }$\mathbf{5\times 5\,\text{cm}^2}$: Fish were presented with darkness ($0\,\text{lm/m}^2$) in a $5\times 5\,\text{cm}^2$ squared arena with $3\,\text{mm}$ of depth. 

{\bf Expanding Spot }$\mathbf{5\times 5\,\text{cm}^2}$: An expanding dark spot at different speeds (0.25, 0.5, 1, 1.5, 2.0, 2.5 cm/s) and different orientations ($0^\circ$, $90^\circ$, $180^\circ$, $270^\circ$) was presented in closed loop 4 cm away from the larva. Stimuli were randomized and presented every 2 min. This assay was done in a $5\times 5\,\text{cm}^2$squared arena with $0.3\,\text{cm}$ of depth.

{\bf Dark Transitions }$\mathbf{5\times 5\,\text{cm}^2}$: The spontaneous swimming with light transitions assay was based on \cite{burgess2007modulation}. Fish were left in the dark for 30 minutes and then presented with uniform light at different intensities (0, 12, 44, 104, 232, 447, 790, 1890, 4700 $\,\text{lm/m}^2$) for 3 min in a $5\times 5\,\text{cm}^2$ arena with $3\,\text{mm}$ of depth.  

{\bf Phototaxis }$\mathbf{5\times 5\,\text{cm}^2}$: The phototaxis assay is based on \cite{huang2013spinal} and performed in $5\times 5\,\text{cm}^2$ arena with $3\,\text{mm}$ of depth. The stimulus consists of a uniform brightness of varying intensity (100, 410, 780, 1250 $\,\text{lm/m}^2$) on one side of the fish and darkness on the other. The stimulus was in closed loop with the larva for a duration of $12\,\text{s}$. 

{\bf Forward Optomotor response }$\mathbf{1\times 5\,\text{cm}^2}$: The forward optomotor response assay was performed as described in \cite{severi2014neural}, but using a $1\times 5\,\text{cm}^2$ with $8\,\text{mm}$ depth arena. Drifting gratings with a spatial period of $1\,\text{cm}$ and ten different speeds (0, 2.5, 5, 7.5, 10, 15, 20, 30, 40, 50 mm/s) were presented from below when the larva is at the extremities of the arena. Trials end when fish reach the opposite end of the arena. After, there is 5s of intertrial interval of homogenous light ($1000\,\text{lm/m}^2$) and a new trial starts with gratings in the opposite direction. In trials that larvae are not able to reach the opposite end of the arena ($> 30\,\text{s}$) a $10\,\text{mm/s}$ grating was displayed until it swam the remaining distance. 

{\bf Directional Optomotor response }$\mathbf{5\times 5\,\text{cm}^2}$: The directional optomotor response assay was performed as in \cite{orger2008control}. A $\mathbf{5\times 5\,\text{cm}^2}$ arena with $3\,\text{mm}$ of depth was used. Drifting gratings with a spatial period of $1\,\text{cm}$, moving at $10\,\text{mm/s}$, and from 24 different orientations ($15^\circ$ apart) were presented from below to the larva and in closed loop with its orientation. Trials last 10s and start when the larva wass in the center of the arena. During the inter-trial interval, that would last at least 5s, circular converging gratings were projected to drive the larva to the center of the arena. 

{\bf High Lux light-dark transitions }$\mathbf{5\times 5\,\text{cm}^2}$: Fish were exposed to alternating 3 min periods of high illuminance light ($5000\,\text{lm/m}^2$) and darkness in an $5\times 5\,\text{cm}^2$ arena with $3\,\text{mm}$ of depth.  

{\bf Prey Capture }$\mathbf{2.5\times 2.5\,\text{cm}^2}$: All prey capture datasets were performed in $2.5\times 2.5\,\text{cm}^2$ arenas with $3\,\text{mm}$ depth and larvae were illuminated from above ($1000\,\text{lm/m}^2$). Fish were fed with 50-100 paramecia (\emph{Paramecium caudatum}) or rotifers (\emph{Brachionus plicatilis}) and were allowed to hunt for 1-2h. A subset of fish were never fed until the assay, while others are raised, starting at 3 days post fertilization, with the type of prey that is used in the assay. To ensure that these fish were not satiated they are starved for at least 2 hours before the assay. 

{\bf Freely exploring in the light, raised with rotifers }$\mathbf{2.5\times 2.5\,\text{cm}^2}$: Fish were fed with 50-100 rotifers starting at 3 days post fertilization and then placed in $2.5\times 2.5\,\text{cm}^2$ arenas with $3\,\text{mm}$ depth and illuminated from above ($1000\,\text{lm/m}^2$) to freely swim. To ensure that these fish were not satiated they were starved for at least 2 hours before the assay.

\section*{Code and Data Availability}
The code used to generate the results and figures in this work is freely available on Github: \url{https://github.com/GautamSridhar/Multiscale_phenotypic_variation}. The data and the analysis results are freely available on Kaggle:
\url{https://doi.org/10.34740/kaggle/dsv/20089164}

\section*{Acknowledgements}
We would like to thank the Wyart lab ``\href{https://wyartlab.org; https://parisbraininstitute.org/paris-brain-institute-research-teams/sibbil-navigation-sensorimotor-integration-brain-body-integration-lab}{SIBBIL}'' for their inputs. We would also like to thank Thierry Emonet, David Jordan and João Costa, for constructive discussions that helped shape the manuscript. We thank the 2020 Fondation Bettencourt-Schueller (FBS-don-0031) award (``Identity and organization or neuronal networks controlling exploration''), the 2020 Prize Equipe ``Fondation pour la Recherche Médicale'' (FRM-EQU202003010612) ‘Neuronal circuits underlying navigation: from genes to behavioral models’, the 2020 European Research Council Consolidator grant no. 101002870 (2021–2026) ``Exploratome: Circuit mechanisms underlying sensory-evoked navigation'' awarded to CW., the European Union’s Horizon 2020 research and innovation program under a Marie Skłodowska-Curie grant no. 813457 ZENITH awarded to CW. that covered the salary of GS. The project benefited as well from the support of the Agence Nationale pour la Recherche (ANR) ANR-22-CE37-0023 LOCOCONNECT, ANR-23-CE16-0017-02 RocSMAP, ANR-24-CE16-7992 CIRCOLOCO, ANR-21-CE14-0042 MOTOMYO and ANR-21-CE13-0008 ASCENTS. GS was also partially supported by funds from OIST Graduate University. ACC acknowledges support from the Agence Nationale pour la Recherche (ANR) with the project ANR-25-CE02-1309. This research was supported in part by the Richard Mille Foundation and the National Science Foundation (NSF) grant PHY-2309135 to the Kavli Institute for Theoretical Physics (KITP) and the Gordon and Betty Moore Foundation Grant No. 2919.02.

\noindent{\bf Author contributions:} G.S., C.W. and A.C.C conceived the study. G.S. and A.C.C developed the methods and analysis code, prepared figures and performed mathematical modeling. G.S. and A.C.C. wrote the original draft. All authors reviewed and edited the manuscript. C.W. and A.C.C. supervised the research and secured funding.

\noindent{\bf Competing interests:} The authors declare no competing interests.

\noindent{\bf Ethics declaration:} not applicable.

\bibliography{apssamp}

@article{laskowski2022emergence,
  title={The emergence and development of behavioral individuality in clonal fish},
  author={Laskowski, Kate L and Bierbach, David and Jolles, Jolle W and Doran, Carolina and Wolf, Max},
  journal={Nature Communications},
  volume={13},
  number={1},
  pages={6419},
  year={2022},
  publisher={Nature Publishing Group UK London}
}

@article{sheftel2013geometry,
  title={The geometry of the p areto front in biological phenotype space},
  author={Sheftel, Hila and Shoval, Oren and Mayo, Avi and Alon, Uri},
  journal={Ecology and evolution},
  volume={3},
  number={6},
  pages={1471--1483},
  year={2013},
  publisher={Wiley Online Library}
}

@article{kollmann2005design,
  title={Design principles of a bacterial signalling network},
  author={Kollmann, Markus and L{\o}vdok, Linda and Bartholom{\'e}, Kilian and Timmer, Jens and Sourjik, Victor},
  journal={Nature},
  volume={438},
  number={7067},
  pages={504--507},
  year={2005},
  publisher={Nature Publishing Group UK London}
}

@article{lovdok2009role,
  title={Role of translational coupling in robustness of bacterial chemotaxis pathway},
  author={L{\o}vdok, Linda and Bentele, Kajetan and Vladimirov, Nikita and M{\"u}ller, Anette and Pop, Ferencz S and Lebiedz, Dirk and Kollmann, Markus and Sourjik, Victor},
  journal={PLoS biology},
  volume={7},
  number={8},
  pages={e1000171},
  year={2009},
  publisher={Public Library of Science San Francisco, USA}
}

@article{reuter19gpcca,
    author = {Reuter,Bernhard and Fackeldey,Konstantin and Weber,Marcus },
    title = {Generalized Markov modeling of nonreversible molecular kinetics},
    journal = {The Journal of Chemical Physics},
    volume = {150},
    number = {17},
    pages = {174103},
    year = {2019},
    doi = {10.1063/1.5064530},
}

@article{laskowski2022consistent,
  title={Consistent individual behavioral variation: what do we know and where are we going?},
  author={Laskowski, Kate L and Chang, Chia-Chen and Sheehy, Kirsten and Agui{\~n}aga, Jonathan},
  journal={Annual Review of Ecology, Evolution, and Systematics},
  volume={53},
  number={1},
  pages={161--182},
  year={2022},
  publisher={Annual Reviews}
}

@article{plevska2021nongenetic,
  title={Nongenetic individuality, changeability, and inheritance in bacterial behavior},
  author={Ple{\v{s}}ka, Maro{\v{s}} and Jordan, David and Frentz, Zak and Xue, BingKan and Leibler, Stanislas},
  journal={Proceedings of the National Academy of Sciences},
  volume={118},
  number={13},
  pages={e2023322118},
  year={2021},
  publisher={National Academy of Sciences}
}

@article{prinz2011markov,
  title={Markov models of molecular kinetics: Generation and validation},
  author={Prinz, Jan-Hendrik and Wu, Hao and Sarich, Marco and Keller, Bettina and Senne, Martin and Held, Martin and Chodera, John D and Sch{\"u}tte, Christof and No{\'e}, Frank},
  journal={The Journal of chemical physics},
  volume={134},
  number={17},
  year={2011},
  publisher={AIP Publishing}
}

@article{chang1970practical,
  title={A practical difference scheme for Fokker-Planck equations},
  author={Chang, JS and Cooper, G},
  journal={Journal of Computational Physics},
  volume={6},
  number={1},
  pages={1--16},
  year={1970},
  publisher={Elsevier}
}

@article{lorenz1963deterministic,
      author = "Edward N.  Lorenz",
      title = "Deterministic Nonperiodic Flow",
      journal = "Journal of Atmospheric Sciences",
      year = "1963",
      publisher = "American Meteorological Society",
      address = "Boston MA, USA",
      volume = "20",
      number = "2",
      doi = "10.1175/1520-0469(1963)020<0130:DNF>2.0.CO;2",
      pages=      "130 - 141",
}

@article{frederickson1983liapunov,
  title={The Liapunov dimension of strange attractors},
  author={Frederickson, Paul and Kaplan, James L and Yorke, Ellen D and Yorke, James A},
  journal={Journal of differential equations},
  volume={49},
  number={2},
  pages={185--207},
  year={1983},
  publisher={Elsevier}
}

@book{sparrow2012lorenz,
  title={The Lorenz equations: bifurcations, chaos, and strange attractors},
  author={Sparrow, Colin},
  year={2012},
  publisher={Springer Science \& Business Media}
}

@inproceedings{
kamiya2024koopman,
title={Koopman Operator Based Dynamical Similarity Analysis for Data-driven Quantification of Distance between Dynamics},
author={Shunsuke Kamiya and Jun Kitazono and Masafumi Oizumi},
booktitle={ICLR 2024 Workshop on Representational Alignment},
year={2024},
url={https://openreview.net/forum?id=cfO9qeHrKU}
}

@article{costa2023maximally,
    author = {Costa, Antonio C. and Ahamed, Tosif and Jordan, David and Stephens, Greg J.},
    title = "{Maximally predictive states: From partial observations to long timescales}",
    journal = {Chaos: An Interdisciplinary Journal of Nonlinear Science},
    volume = {33},
    number = {2},
    pages = {023136},
    year = {2023},
    month = {02},
    issn = {1054-1500},
    doi = {10.1063/5.0129398},
    url = {https://doi.org/10.1063/5.0129398},
}

@article{marques2018structure,
  title={Structure of the zebrafish locomotor repertoire revealed with unsupervised behavioral clustering},
  author={Marques, Jo{\~a}o C and Lackner, Simone and F{\'e}lix, Rita and Orger, Michael B},
  journal={Current Biology},
  volume={28},
  number={2},
  pages={181--195},
  year={2018},
  publisher={Elsevier}
}

@article{severi2014neural,
  title={Neural control and modulation of swimming speed in the larval zebrafish},
  author={Severi, Kristen E and Portugues, Ruben and Marques, Jo{\~a}o C and O’Malley, Donald M and Orger, Michael B and Engert, Florian},
  journal={Neuron},
  volume={83},
  number={3},
  pages={692--707},
  year={2014},
  publisher={Elsevier}
}

@article{froyland2003detecting,
  title={Detecting and locating near-optimal almost-invariant sets and cycles},
  author={Froyland, Gary and Dellnitz, Michael},
  journal={SIAM Journal on Scientific Computing},
  volume={24},
  number={6},
  pages={1839--1863},
  year={2003},
  publisher={SIAM}
}

@article{froyland2005statistically,
  title={Statistically optimal almost-invariant sets},
  author={Froyland, Gary},
  journal={Physica D: Nonlinear Phenomena},
  volume={200},
  number={3-4},
  pages={205--219},
  year={2005},
  publisher={Elsevier}
}

@article{zelnik2004self,
  title={Self-tuning spectral clustering},
  author={Zelnik-Manor, Lihi and Perona, Pietro},
  journal={Advances in neural information processing systems},
  volume={17},
  year={2004}
}

@book{chavel1984eigenvalues,
  title={Eigenvalues in Riemannian Geometry},
  author={Chavel, I. and Randol, B. and Dodziuk, J.},
  isbn={9780080874340},
  series={Pure and Applied Mathematics},
  url={https://books.google.fr/books?id=0v1VfTWuKGgC},
  year={1984},
  publisher={Academic Press}
}

@book{petersen2006riemannian,
  title={Riemannian geometry},
  author={Petersen, Peter},
  year={2006},
  publisher={Springer}
}

@ARTICLE{2020SciPy-NMeth,
  author  = {Virtanen, Pauli and Gommers, Ralf and Oliphant, Travis E. and
            Haberland, Matt and Reddy, Tyler and Cournapeau, David and
            Burovski, Evgeni and Peterson, Pearu and Weckesser, Warren and
            Bright, Jonathan and {van der Walt}, St{\'e}fan J. and
            Brett, Matthew and Wilson, Joshua and Millman, K. Jarrod and
            Mayorov, Nikolay and Nelson, Andrew R. J. and Jones, Eric and
            Kern, Robert and Larson, Eric and Carey, C J and
            Polat, {\.I}lhan and Feng, Yu and Moore, Eric W. and
            {VanderPlas}, Jake and Laxalde, Denis and Perktold, Josef and
            Cimrman, Robert and Henriksen, Ian and Quintero, E. A. and
            Harris, Charles R. and Archibald, Anne M. and
            Ribeiro, Ant{\^o}nio H. and Pedregosa, Fabian and
            {van Mulbregt}, Paul and {SciPy 1.0 Contributors}},
  title   = {{{SciPy} 1.0: Fundamental Algorithms for Scientific
            Computing in Python}},
  journal = {Nature Methods},
  year    = {2020},
  volume  = {17},
  pages   = {261--272},
  adsurl  = {https://rdcu.be/b08Wh},
  doi     = {10.1038/s41592-019-0686-2},
}

@book{bollt2013applied,
  title={Applied and computational measurable dynamics},
  author={Bollt, Erik M and Santitissadeekorn, Naratip},
  year={2013},
  publisher={SIAM}
}

@book{berg2004coli,
  title={E. coli in Motion},
  author={Berg, Howard C},
  year={2004},
  publisher={Springer}
}

@article{burgess2007modulation,
  title={Modulation of locomotor activity in larval zebrafish during light adaptation},
  author={Burgess, Harold A and Granato, Michael},
  journal={Journal of Experimental Biology},
  volume={210},
  number={14},
  pages={2526--2539},
  year={2007},
  publisher={Company of Biologists}
}

@article{noe2015kinetic,
  title={Kinetic distance and kinetic maps from molecular dynamics simulation},
  author={No{\'e}, Frank and Clementi, Cecilia},
  journal={Journal of chemical theory and computation},
  volume={11},
  number={10},
  pages={5002--5011},
  year={2015},
  publisher={ACS Publications}
}

@article{mcelligott2005prey,
  title={Prey tracking by larval zebrafish: axial kinematics and visual control},
  author={McElligott, Melissa B and O’Malley, Donald M},
  journal={Brain Behavior and Evolution},
  volume={66},
  number={3},
  pages={177--196},
  year={2005},
  publisher={S. Karger AG}
}

@article{bianco2011prey,
  title={Prey capture behavior evoked by simple visual stimuli in larval zebrafish},
  author={Bianco, Isaac H and Kampff, Adam R and Engert, Florian},
  journal={Frontiers in systems neuroscience},
  volume={5},
  pages={101},
  year={2011},
  publisher={Frontiers Research Foundation}
}

@article{hernandez2021framework,
  title={A framework for studying behavioral evolution by reconstructing ancestral repertoires},
  author={Hern{\'a}ndez, Dami{\'a}n G and Rivera, Catalina and Cande, Jessica and Zhou, Baohua and Stern, David L and Berman, Gordon J},
  journal={Elife},
  volume={10},
  pages={e61806},
  year={2021},
  publisher={eLife Sciences Publications Limited}
}

@article{roth2003optimal,
  title={Optimal cluster preserving embedding of nonmetric proximity data},
  author={Roth, Volker and Laub, Julian and Kawanabe, Motoaki and Buhmann, Joachim M},
  journal={IEEE Transactions on Pattern Analysis and Machine Intelligence},
  volume={25},
  number={12},
  pages={1540--1551},
  year={2003},
  publisher={IEEE}
}

@article{costa2024markovian,
author = {Antonio C. Costa  and Tosif Ahamed  and David Jordan  and Greg J. Stephens },
title = {A Markovian dynamics for \textit{Caenorhabditis elegans} behavior across scales},
journal = {Proceedings of the National Academy of Sciences},
volume = {121},
number = {32},
pages = {e2318805121},
year = {2024},
doi = {10.1073/pnas.2318805121},
URL = {https://www.pnas.org/doi/abs/10.1073/pnas.2318805121},
eprint = {https://www.pnas.org/doi/pdf/10.1073/pnas.2318805121}}

@article{lagogiannis2020learning,
  title={Learning steers the ontogeny of an efficient hunting sequence in zebrafish larvae},
  author={Lagogiannis, Konstantinos and Diana, Giovanni and Meyer, Martin P},
  journal={Elife},
  volume={9},
  pages={e55119},
  year={2020},
  publisher={eLife Sciences Publications, Ltd}
}

@article{oldfield2020experience,
  title={Experience, circuit dynamics, and forebrain recruitment in larval zebrafish prey capture},
  author={Oldfield, Claire S and Grossrubatscher, Irene and Chavez, Mario and Hoagland, Adam and Huth, Alex R and Carroll, Elizabeth C and Prendergast, Andrew and Qu, Tony and Gallant, Jack L and Wyart, Claire and others},
  journal={Elife},
  volume={9},
  pages={e56619},
  year={2020},
  publisher={eLife Sciences Publications, Ltd}
}

@article{sih2004behavioral,
  title={Behavioral syndromes: an ecological and evolutionary overview},
  author={Sih, Andrew and Bell, Alison and Johnson, J Chadwick},
  journal={Trends in ecology \& evolution},
  volume={19},
  number={7},
  pages={372--378},
  year={2004},
  publisher={Elsevier}
}

@article{berman2018measuring,
  title={Measuring behavior across scales},
  author={Berman, Gordon J},
  journal={BMC biology},
  volume={16},
  pages={1--11},
  year={2018},
  publisher={Springer}
}

@article{costa2024fluctuating,
  title={Fluctuating landscapes and heavy tails in animal behavior},
  author={Costa, Antonio Carlos and Sridhar, Gautam and Wyart, Claire and Vergassola, Massimo},
  journal={PRX Life},
  volume={2},
  number={2},
  pages={023001},
  year={2024},
  publisher={APS}
}

@article{brown2018ethology,
  title={Ethology as a physical science},
  author={Brown, Andr{\'e} EX and De Bivort, Benjamin},
  journal={Nature Physics},
  volume={14},
  number={7},
  pages={653--657},
  year={2018},
  publisher={Nature Publishing Group UK London}
}

@article{orger2008control,
  title={Control of visually guided behavior by distinct populations of spinal projection neurons},
  author={Orger, Michael B and Kampff, Adam R and Severi, Kristen E and Bollmann, Johann H and Engert, Florian},
  journal={Nature neuroscience},
  volume={11},
  number={3},
  pages={327--333},
  year={2008},
  publisher={Nature Publishing Group US New York}
}

@article{marder2023individual,
  title={Individual variability, statistics, and the resilience of nervous systems of crabs and humans to temperature and other perturbations},
  author={Marder, Eve},
  journal={Eneuro},
  volume={10},
  number={11},
  year={2023},
  publisher={Society for Neuroscience}
}

@article{deBivort2022precise,
  title={Precise quantification of behavioral individuality from 80 million decisions across 183,000 flies},
  author={de Bivort, Benjamin and Buchanan, Sean and Skutt-Kakaria, Kyobi and Gajda, Erika and Ayroles, Julien and O’Leary, Chelsea and Reimers, Pablo and Akhund-Zade, Jamilla and Senft, Rebecca and Maloney, Ryan and others},
  journal={Frontiers in Behavioral Neuroscience},
  volume={16},
  pages={836626},
  year={2022},
  publisher={Frontiers}
}

@article{golowasch2002failure,
  title={Failure of averaging in the construction of a conductance-based neuron model},
  author={Golowasch, Jorge and Goldman, Mark S and Abbott, LF and Marder, Eve},
  journal={Journal of neurophysiology},
  volume={87},
  number={2},
  pages={1129--1131},
  year={2002},
  publisher={American Physiological Society Bethesda, MD}
}

@article{pantoja2016neuromodulatory,
  title={Neuromodulatory regulation of behavioral individuality in zebrafish},
  author={Pantoja, Carlos and Hoagland, Adam and Carroll, Elizabeth C and Karalis, Vasiliki and Conner, Alden and Isacoff, Ehud Y},
  journal={Neuron},
  volume={91},
  number={3},
  pages={587--601},
  year={2016},
  publisher={Elsevier}
}

@article{werkhoven2021structure,
  title={The structure of behavioral variation within a genotype},
  author={Werkhoven, Zachary and Bravin, Alyssa and Skutt-Kakaria, Kyobi and Reimers, Pablo and Pallares, Luisa F and Ayroles, Julien and De Bivort, Benjamin L},
  journal={Elife},
  volume={10},
  pages={e64988},
  year={2021},
  publisher={eLife Sciences Publications Limited}
}

@article{datta2019computational,
  title={Computational neuroethology: a call to action},
  author={Datta, Sandeep Robert and Anderson, David J and Branson, Kristin and Perona, Pietro and Leifer, Andrew},
  journal={Neuron},
  volume={104},
  number={1},
  pages={11--24},
  year={2019},
  publisher={Elsevier}
}

@article{huang2013spinal,
  title={Spinal projection neurons control turning behaviors in zebrafish},
  author={Huang, Kuo-Hua and Ahrens, Misha B and Dunn, Timothy W and Engert, Florian},
  journal={Current Biology},
  volume={23},
  number={16},
  pages={1566--1573},
  year={2013},
  publisher={Elsevier}
}

@article{dufour2016direct,
  title={Direct correlation between motile behavior and protein abundance in single cells},
  author={Dufour, Yann S and Gillet, S{\'e}bastien and Frankel, Nicholas W and Weibel, Douglas B and Emonet, Thierry},
  journal={PLoS computational biology},
  volume={12},
  number={9},
  pages={e1005041},
  year={2016},
  publisher={Public Library of Science San Francisco, CA USA}
}

@article{sridhar2024uncovering,
  title={Uncovering multiscale structure in the variability of larval zebrafish navigation},
  author={Sridhar, Gautam and Vergassola, Massimo and Marques, Jo{\~a}o C and Orger, Michael B and Costa, Antonio Carlos and Wyart, Claire},
  journal={Proceedings of the National Academy of Sciences},
  volume={121},
  number={47},
  pages={e2410254121},
  year={2024},
  publisher={National Academy of Sciences}
}

@article{coifman2006diffusion,
  title={Diffusion maps},
  author={Coifman, Ronald R and Lafon, St{\'e}phane},
  journal={Applied and computational harmonic analysis},
  volume={21},
  number={1},
  pages={5--30},
  year={2006},
  publisher={Elsevier}
}

@article{Wadhams2004,
  title = {Making sense of it all: bacterial chemotaxis},
  volume = {5},
  ISSN = {1471-0080},
  url = {http://dx.doi.org/10.1038/nrm1524},
  DOI = {10.1038/nrm1524},
  number = {12},
  journal = {Nature Reviews Molecular Cell Biology},
  publisher = {Springer Science and Business Media LLC},
  author = {Wadhams,  George H. and Armitage,  Judith P.},
  year = {2004},
  month = dec,
  pages = {1024–1037}
}

@article{Spudich1976,
  title = {Non-genetic individuality: chance in the single cell},
  volume = {262},
  ISSN = {1476-4687},
  url = {http://dx.doi.org/10.1038/262467a0},
  DOI = {10.1038/262467a0},
  number = {5568},
  journal = {Nature},
  publisher = {Springer Science and Business Media LLC},
  author = {Spudich,  John L. and Koshland,  D. E.},
  year = {1976},
  month = aug,
  pages = {467–471}
}

@article{Park2010,
  title = {Interdependence of behavioural variability and response to small stimuli in bacteria},
  volume = {468},
  ISSN = {1476-4687},
  url = {http://dx.doi.org/10.1038/nature09551},
  DOI = {10.1038/nature09551},
  number = {7325},
  journal = {Nature},
  publisher = {Springer Science and Business Media LLC},
  author = {Park,  Heungwon and Pontius,  William and Guet,  Calin C. and Marko,  John F. and Emonet,  Thierry and Cluzel,  Philippe},
  year = {2010},
  month = nov,
  pages = {819–823}
}

@incollection{sourjik2007vivo,
  title={In vivo measurement by FRET of pathway activity in bacterial chemotaxis},
  author={Sourjik, Victor and Vaknin, Ady and Shimizu, Thomas S and Berg, Howard C},
  booktitle={Methods in enzymology},
  volume={423},
  pages={365--391},
  year={2007},
  publisher={Elsevier}
}

@article{tao2019statistical,
  title={Statistical structure of locomotion and its modulation by odors},
  author={Tao, Liangyu and Ozarkar, Siddhi and Beck, Jeffrey M and Bhandawat, Vikas},
  journal={Elife},
  volume={8},
  pages={e41235},
  year={2019},
  publisher={eLife Sciences Publications, Ltd}
}

@article{transtrum2015perspective,
  title={Perspective: Sloppiness and emergent theories in physics, biology, and beyond},
  author={Transtrum, Mark K and Machta, Benjamin B and Brown, Kevin S and Daniels, Bryan C and Myers, Christopher R and Sethna, James P},
  journal={The Journal of chemical physics},
  volume={143},
  number={1},
  year={2015},
  publisher={AIP Publishing}
}

@article{ostrow2023beyond,
  title={Beyond geometry: Comparing the temporal structure of computation in neural circuits with dynamical similarity analysis},
  author={Ostrow, Mitchell and Eisen, Adam and Kozachkov, Leo and Fiete, Ila},
  journal={Advances in Neural Information Processing Systems},
  volume={36},
  pages={33824--33837},
  year={2023}
}

@article{huang2025inputdsa,
  title={Inputdsa: Demixing then comparing recurrent and externally driven dynamics},
  author={Huang, Ann and Ostrow, Mitchell and Singh, Satpreet H and Kozachkov, Leo and Fiete, Ila and Rajan, Kanaka},
  journal={arXiv preprint arXiv:2510.25943},
  year={2025}
}

@inproceedings{zhang2025koopstd,
  title={KoopSTD: Reliable Similarity Analysis between Dynamical Systems via Approximating Koopman Spectrum with Timescale Decoupling},
  author={Zhang, Shimin and Ye, Ziyuan and Yan, Yinsong and Song, Zeyang and Wu, Yujie and Wu, Jibin},
  booktitle={Forty-second International Conference on Machine Learning},
  year={2025}
}

@book{kutz2016dynamic,
  title={Dynamic mode decomposition: data-driven modeling of complex systems},
  author={Kutz, J Nathan and Brunton, Steven L and Brunton, Bingni W and Proctor, Joshua L},
  year={2016},
  publisher={SIAM}
}

@article{van2008visualizing,
  title={Visualizing data using t-SNE.},
  author={Van der Maaten, Laurens and Hinton, Geoffrey},
  journal={Journal of machine learning research},
  volume={9},
  number={11},
  year={2008}
}

@article{roweis2000nonlinear,
  title={Nonlinear dimensionality reduction by locally linear embedding},
  author={Roweis, Sam T and Saul, Lawrence K},
  journal={science},
  volume={290},
  number={5500},
  pages={2323--2326},
  year={2000},
  publisher={American Association for the Advancement of Science}
}

@article{mcinnes2018umap,
  title={Umap: Uniform manifold approximation and projection for dimension reduction},
  author={McInnes, Leland and Healy, John and Melville, James},
  journal={arXiv preprint arXiv:1802.03426},
  year={2018}
}

@article{germain2025spectral,
  title={A Spectral-Grassmann Wasserstein metric for operator representations of dynamical systems},
  author={Germain, Thibaut and Flamary, R{\'e}mi and Kostic, Vladimir R and Lounici, Karim},
  journal={arXiv preprint arXiv:2509.24920},
  year={2025}
}

@article{cohen2023schrodinger,
  title={Schr{\"o}dinger dynamics and berry phase of undulatory locomotion},
  author={Cohen, Alexander E and Hastewell, Alasdair D and Pradhan, Sreeparna and Flavell, Steven W and Dunkel, J{\"o}rn},
  journal={Physical Review Letters},
  volume={130},
  number={25},
  pages={258402},
  year={2023},
  publisher={APS}
}

@article{frankel2014adaptability,
  title={Adaptability of non-genetic diversity in bacterial chemotaxis},
  author={Frankel, Nicholas W and Pontius, William and Dufour, Yann S and Long, Junjiajia and Hernandez-Nunez, Luis and Emonet, Thierry},
  journal={Elife},
  volume={3},
  pages={e03526},
  year={2014},
  publisher={eLife Sciences Publications, Ltd}
}

@article{sato2020evolutionary,
  title={Evolutionary dimension reduction in phenotypic space},
  author={Sato, Takuya U and Kaneko, Kunihiko},
  journal={Physical Review Research},
  volume={2},
  number={1},
  pages={013197},
  year={2020},
  publisher={APS}
}

@article{kaneko2024constructing,
  title={Constructing universal phenomenology for biological cellular systems: an idiosyncratic review on evolutionary dimensional reduction},
  author={Kaneko, Kunihiko},
  journal={Journal of Statistical Mechanics: Theory and Experiment},
  volume={2024},
  number={2},
  pages={024002},
  year={2024},
  publisher={IOP Publishing}
}

@article{chihoub2025evolution,
  title={The evolution of robustness and fragility during long-term bacterial adaptation},
  author={Chihoub, Doha and Pintard, Coralie and Lenski, Richard E and Tenaillon, Olivier and Couce, Alejandro},
  journal={Proceedings of the National Academy of Sciences},
  volume={122},
  number={16},
  pages={e2501901122},
  year={2025},
  publisher={National Academy of Sciences}
}

@article{romeo2025characterizing,
  title={Characterizing nonlinear dynamics by contrastive cartography},
  author={Romeo, Nicolas and Chi, Chris and Dinner, Aaron R and Jerison, Elizabeth R},
  journal={PRX Life},
  volume={3},
  number={3},
  pages={033022},
  year={2025},
  publisher={APS}
}

@article{loscher2024mice,
  title={Of Mice and Men: The Inter-individual Variability of the Brain's Response to Drugs},
  author={L{\"o}scher, Wolfgang},
  journal={Eneuro},
  volume={11},
  number={2},
  year={2024},
  publisher={Society for Neuroscience}
}

@article{shoval2012evolutionary,
  title={Evolutionary trade-offs, Pareto optimality, and the geometry of phenotype space},
  author={Shoval, Oren and Sheftel, Hila and Shinar, Guy and Hart, Yuval and Ramote, Omer and Mayo, Avi and Dekel, Erez and Kavanagh, Kathryn and Alon, Uri},
  journal={Science},
  volume={336},
  number={6085},
  pages={1157--1160},
  year={2012},
  publisher={American Association for the Advancement of Science}
}

@article{huttegger2011invariance,
  title={Invariance and meaningfulness in phenotype spaces},
  author={Huttegger, Simon M and Mitteroecker, Philipp},
  journal={Evolutionary Biology},
  volume={38},
  number={3},
  pages={335--351},
  year={2011},
  publisher={Springer}
}

\clearpage
\onecolumngrid

\section*{Supplementary Information}

\setcounter{figure}{0}
\setcounter{page}{1}
\setcounter{equation}{0}
\setcounter{section}{0}

\makeatletter
\renewcommand{\theequation}{S\arabic{equation}}
\renewcommand{\thefigure}{S\arabic{figure}}
\renewcommand{\thepage}{S\arabic{page}}
\renewcommand{\thesection}{S\arabic{section}}

\begin{figure*}[ht!]
    \includegraphics{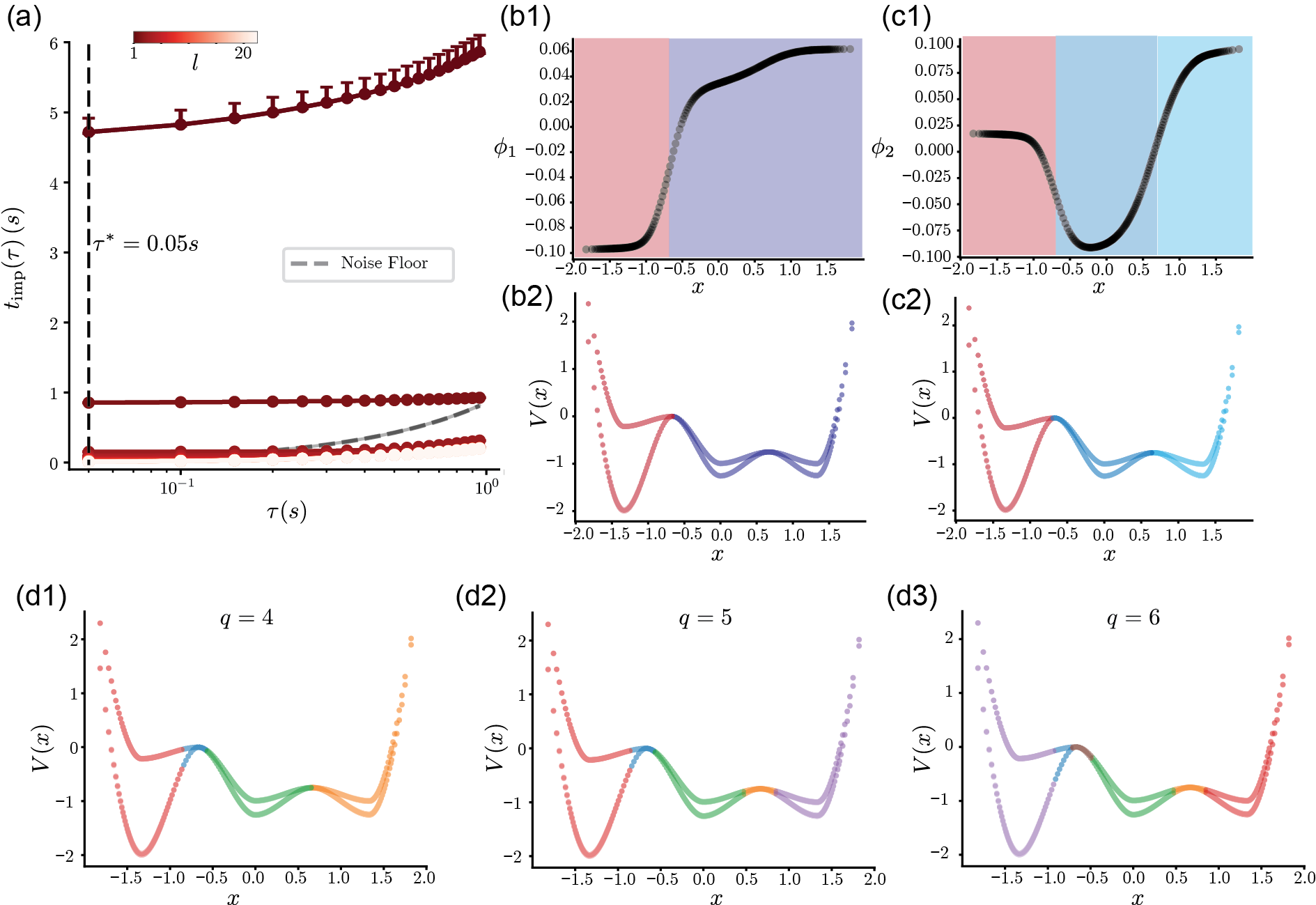}
    \caption{{\bf Details of the estimation of the ensemble transfer operator for the overdamped dynamics on a 3-well potential.} We first partition the state space consisting of all simulations across phenotypes into $N=250$ microstates. (a) Implied timescales $t_\text{imp}(\tau)$. We select $\tau=0.05s$ to construct the ensemble transfer operator $P_\text{ensemble}$ (b1) The first non-trivial eigenvector $\phi_1$ identifies the largest energy barrier in the system, enabling a coarse-graining into 2 states. (c1) The second non-trivial eigenvector $\phi_2$ further identifies the second-largest energy barrier in the system, resulting in a 3-state Markov chain capturing transitions among the metastable wells. (d1-d3) Coarse-graining for $q=4,5,6$. Subsequent coarse-graining splits finescale dynamics in the transition regions while preserving the bulk of the metastable sets.}
    \label{fig:S1_3well_cg}
\end{figure*}

\begin{figure*}[ht!]
    \includegraphics{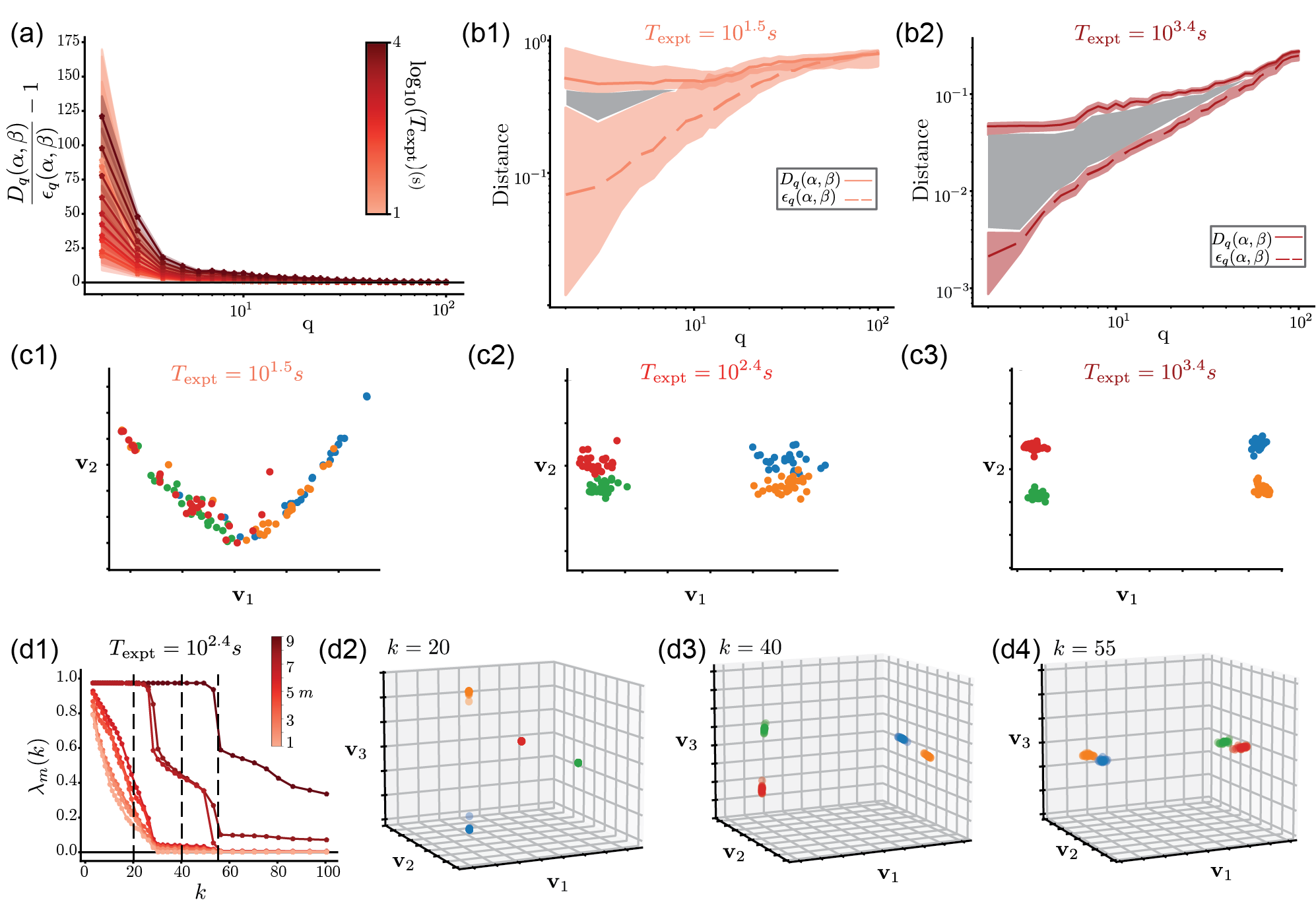}
    \caption{{\bf Dependence of the phenotypic space reconstruction on the simulation time $T_\text{expt}$ and the number of nearest neighbors $k$, for the overdamped dynamics from a 3-well potential.} (a) We plot  $D_q/\varepsilon_q-1$ as a function of $q$, averaged across all observations. The area under each curve represents the multiscale distance metric $D(\alpha,\beta)$. Errorbars represent $95\%$ confidence intervals across simulations. As $T_\text{expt}$ increases, the area under the curve increases, pointing to increasing resolution in the differences between phenotypic groups. (b1,b2)  Two examples of plots of $D_q(\alpha,\beta)$ and $\varepsilon_q(\alpha,\beta)$ at $T_\text{expt} = 10^{1.5}\,\text{s}$ and $T_\text{expt} = 10^{3.4}\,\text{s}$. Shaded areas point to the multiscale distance that increases with $T_\text{expt}$ (c) Projection along the first two diffusion coordinates $(\mathbf{v}_1,\mathbf{v}_2)$ (obtained with $k=55$ nerest neighbors)) for different values of $T_\text{expt}$. (c1) At low $T_\text{expt} = 10^{1.5}\,\text{s}$, the approach cannot distinguish blue from orange or red from green phenotypes, and yields an approximately 1D manifold. (c2) At $T_\text{expt} = 10^{2.4}\,\text{s}$, the first major axes are resolved, but there is still some overlap between phenotypes. (c3) At $T_\text{expt} = 10^{3.4}\,\text{s}$, all phenotypes are resolved completely. (d1-d4) Dependence of the diffusion map eigenvalues and coordinates on the number of nearest neighbors $k$ for $T_\text{expt}=10^{3.4}\,\text{s}$. (d1) Eigenvalues $\lambda_m(k)$ as a function of $k$ for various eigenvalues $m$. We focus on 3 regimes - $k=20$, for which the kernel bandwidth corresponds to within-cluster distances, $k=40$, for which the kernel bandwidth now reaches the nearest cluster, and $k=55$, for which the kernel bandwidth reaches all clusters. We notice a local vs. global information trade-off slowly unraveling as $k$ increases. (d2)  For $k=20$, the information is local, we receive $\lambda_{m\in{1,2,3}} \approx 1$ and separated from the bulk. In this regime, the diffusion map eigenvectors $\mathbf{v}_1,\mathbf{v}_2,\mathbf{v}_3$ separate each cluster directly.(d3) For $k=40$, the global access of variation begin to be resolved. (d4) At $k=55$, the global information capturing 2 major axes of variation becomes clear.}
    \label{fig:S2_3well_Texpt_k}
\end{figure*}

\begin{figure*}[ht!]
    \includegraphics{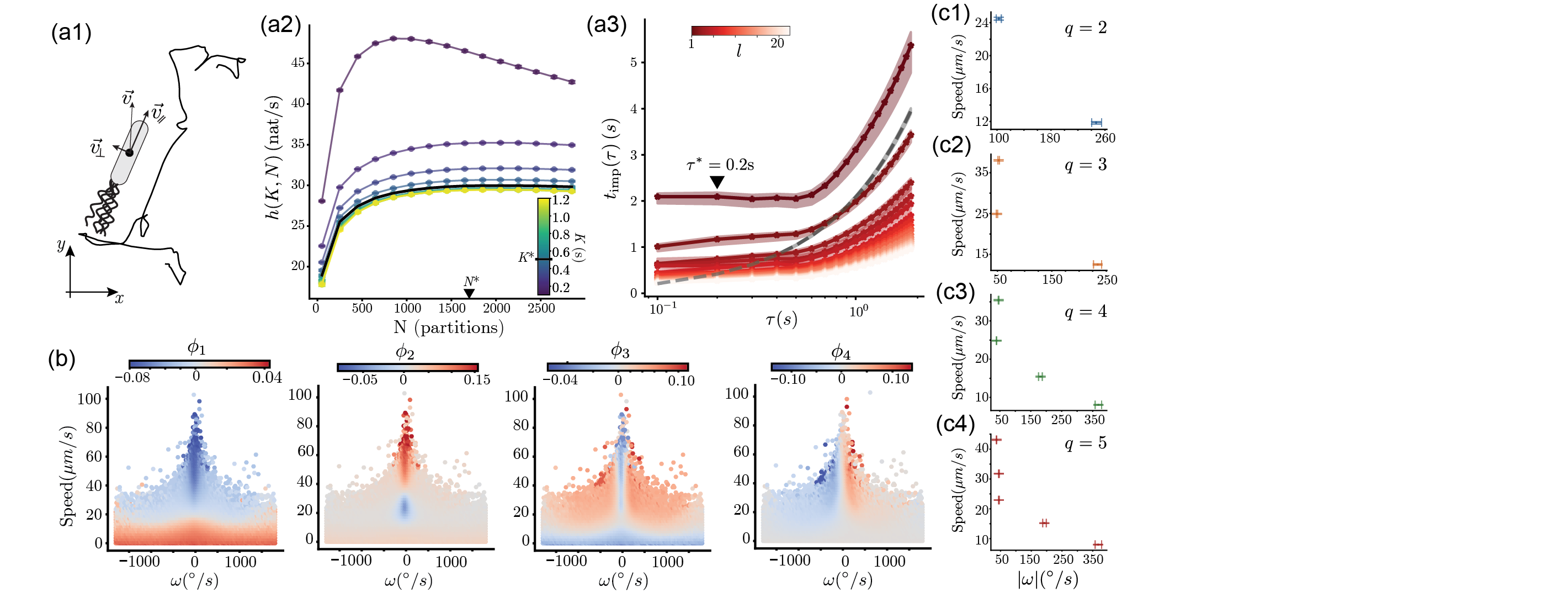}
    \caption{{\bf Details on the transfer operator approximation and coarse-grained dynamics for the bacterial chemotaxis dataset from \cite{dufour2016direct}.} (a1) We begin by estimating the parallel and perpendicular components of the velocity vectors estimated from the centroid trajectories. (a2) To ensure a maximally-predictive Markovian description, we perform a time delay embedding, and choose the number of delays $K$ by minimizing the entropy rate of the resulting Markov chain \cite{costa2023maximally} (see Methods). We select $K^*=0.5\,\text{s}$ to minimize the entropy rate, and $N^*=1700$ microstates to maximize the resolution over the dynamics before finite-size effects impair the estimation of the entropy rate. (a3) Implied timescales $t_\text{imp}(\tau)$. We select $\tau^*=0.2s$ to construct the ensemble transfer operator, in a regime where the inferred timescales are nearly invariant to $\tau$, indicating Markovianity \cite{costa2023maximally}. (b) We report the first $\phi_{l\in[1,4]}$ non-trivial eigenvectors: $\phi_1$ differentiates running and tumbling behavior, $\phi_2$ differentiates an intermediate speed state from running and tumbling, $\phi_3$ differentiates smooth turning behaviors from straight runs and tumbles, and $\phi_4$ differentiates left-biased runs from right-biased runs. (c) Mean speeds and absolute changes in orientation $|\omega|$ for $q\in[2,5]$. (c1) At $q=2$, we identify running and tumbling states. (c2) At $q=3$, the run states splits into fast and slow. (c3-c4) As $q$ increases, we find finer scale partitions of the running and tumbling behavior with varying speeds and reorientation rates.}
    \label{fig:S3_bacterial_dyn}
\end{figure*}

\begin{figure*}[ht!]
    \includegraphics{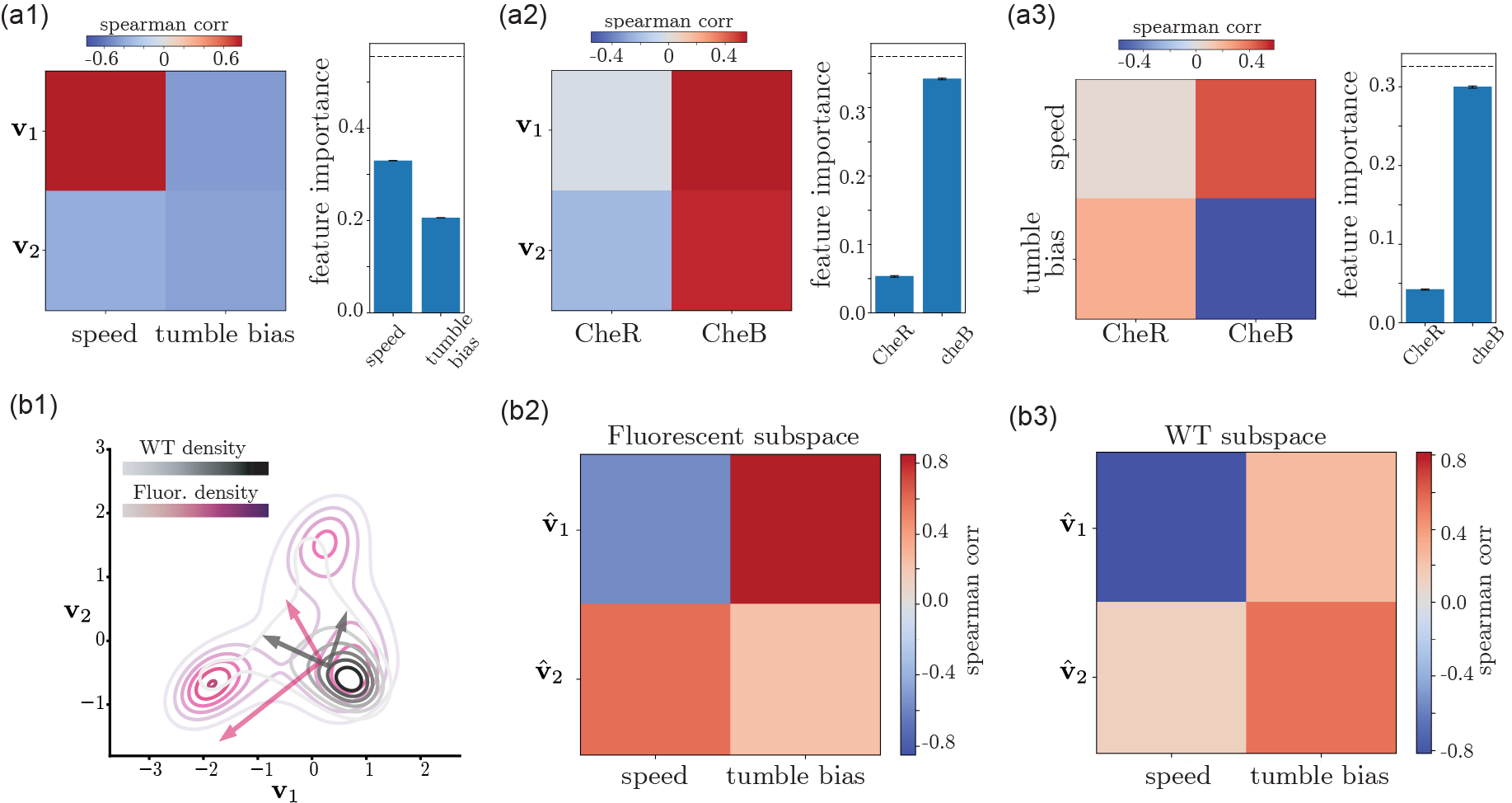}
    \caption{{\bf Phenotypic variation among wild-type and fluorescent bacteria is associated with variation in both run speed and tumble bias, and dominated by fluctuations in CheB.} (a1-left) Pairwise spearman correlation $\rho$ between the projection along the two main modes of variation $\mathbf{v}_1$ and $\mathbf{v}_2$, and the run speed and tumble bias of each individual. The first mode $\mathbf{v}_1$ is mostly correlated with run speeds, $\rho_{\mathbf{v}_1,\text{speed}} = 0.78\,(0.77, 0.79)$, and also with tumble bias, $\rho_{\mathbf{v}_1,\text{tumble bias}} = -0.46\,(-0.48, -0.44)$. The second mode $\mathbf{v}_2$ is nearly equally correlated with tumble bias and speed, $\rho_{\mathbf{v}_2,\text{speed}} = -0.39\,(-0.41, -0.36)$, $\rho_{\mathbf{v}_2,\text{tumble bias}} = -0.44\,(-0.46, -0.42)$. Unless otherwise noted, p-values are negligibly small. Each estimate of the correlation is accompanied by bootstrapped 95\% confidence intervals.
    (a1-right) Importance of either speed or tumble bias in predicting the position of each individual along ($\mathbf{v}_1$,$\mathbf{v}_2$) using linear regression. We measure this by estimating the reduction in the coefficient of determination $R^2$ that results from shuffling either variable. Speed is more important than tumble bias in determining the position of each individual in the reconstructed phenotypic space. Error bars represent 95\% bootstrapped confidence intervals over random reshuffles. The horizontal dashed line represents the $R^2$ of the unshuffled model.
    (a2-left) Same as (a1), but correlating $\mathbf{v}_1$ and $\mathbf{v}_2$ to CheR and CheB number in the mutant strains. Both modes are strongly correlated with CheB numbers, $\rho_{\mathbf{v}_1,\text{cheB}} = 0.55\,(0.52, 0.59),\, \rho_{\mathbf{v}_2,\text{cheB}} =  0.52\,(0.48, 0.55)$, with $\mathbf{v}_2$ also correlated with CheR expression, $\rho_{\mathbf{v}_2,\text{cheR}}= -0.23\,\,(-0.28, -0.18)$, whereas $\mathbf{v}_1$ is uncorrelated with CheR, $\rho_{\mathbf{v}_1,\text{cheR}} =-0.03\,(-0.08, 0.03)$ (p-value=0.33). 
    (a2-right) Same as (a1-right), but for the importance of either CheB or CheR in predicting the position of each individual along ($\mathbf{v}_1$,$\mathbf{v}_2$). The full model yields an  $R^2 = 0.38\,(0.35,0.40)$, with CheB fluctuations capturing most of the predictive power. 
    (a3-left) Same as (a1), but correlating CheR and CheB numbers in the mutant strains to their run speed and tumble bias. Speed is uncorrelated to CheR numbers, $\rho_{\text{speed},\text{cheR}} =-0.03\,(-0.02, 0.09)$ (p-value=0.23), but correlates with CheB, $\rho_{\text{speed},\text{cheB}} =-0.46\,(0.42, 0.50)$. Tumble bias correlates most strongly with CheB, $\rho_{\text{tumble bias},\text{cheB}} =-0.59\,(-0.62, 0.55)$, but also with CheR, $\rho_{\text{tumble bias},\text{cheR}} =0.22\,(0.18, 0.27)$.
    (a3-right) Same as (b1), but for the importance of CheB and CheR in predicting both the run speed and tumble bias of each individual. Once again, CheB is a major determinant of inter-individual variability.
    (b1) Probability distribution function (as a contour plot), and major axis of variation (arrows), for either the mutant data, or the wild-type bacteria. We find that the variation is more anisotropic in the mutant data when compared to wild-type data, and that the main axis of variation from either dataset are rotated by nearly $90^\circ$ clockwise.
    (b2) Same as (a1-left), but estimating the Spearman correlation between the projection along the major axis of variation in the fluorescent dataset, and the speed and tumble bias. The main mode of variation in the fluorescent data $\hat{v}_1$ is mostly correlated with tumble bias, $\rho_{\hat{\mathbf{v}}_1,\text{tumble bias}} = 0.85\,(0.83, 0.86)$, but also speed, $\rho_{\hat{\mathbf{v}}_1,\text{speed}} =-0.57\,(-0.61, -0.52)$. The second mode is mostly correlated with speed, $\rho_{\hat{\mathbf{v}}_2,\text{speed}} =0.59\,(0.55, 0.62)$, and less with tumble bias, $\rho_{\hat{\mathbf{v}}_2,\text{tumble bias}} =0.23\,(0.18, 0.28)$.
    (b3) Same as (b2), but for the dominant modes of variation in the wild-type dataset. We find that $\mathbf{v}_1$ is mostly correlated with speed, $\rho_{\hat{\mathbf{v}}_1,\text{speed}} =-0.81\,(-0.83, -0.80)$, but also somewhat with tumble bias, $\rho_{\hat{\mathbf{v}}_1,\text{tumble bias}} =0.26\,(0.23, 0.28)$. Conversely, $\mathbf{v}_2$ is mostly correlated with tumble bias, $\rho_{\hat{\mathbf{v}}_2,\text{tumble bias}} =0.55\,(0.53, 0.57)$, and less with speed, $\rho_{\hat{\mathbf{v}}_2,\text{speed}} =0.13\,(0.10, 0.16)$. 
    }
    \label{fig:S4_bacteria_corrs}
\end{figure*}

\begin{figure*}[ht!]
    \includegraphics{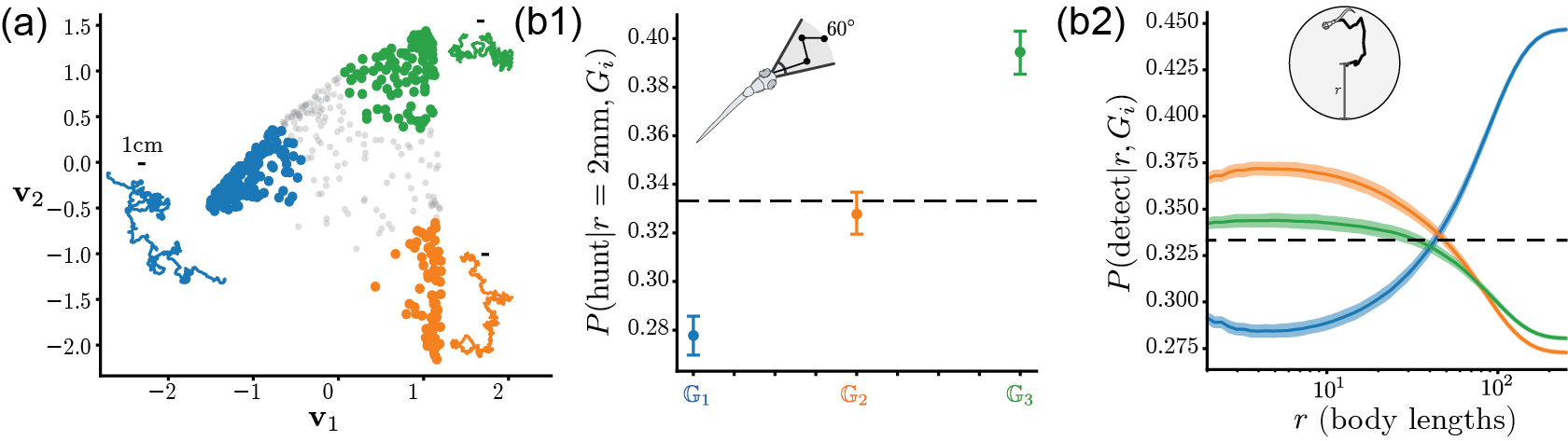}
    \caption{{\bf Analysis of the phenotypic groups from the larval zebrafish dataset.} (a) We utilize fuzzy c-means to define 3 groups at the extremes of the phenotypic space (see Methods). (b1) As in \cite{sridhar2024uncovering}, we simulate boundaryless trajectories from the average transfer operator of the three groups. We calculate the probability of approaching a prey item successfully in each group by measuring the probability of resource collection in a $60^\circ$ field of view from the start point within the first 3 bouts. The dotted line represents the average expected probability among 3 groups of 1/3. We find that $G_3$ is the best phenotype at hunting prey, matching the region with highest eye convergence in the phenotypic space, Fig.\,\ref{fig:5}. (b2) We also calculate the probability of resource collection evenly distributed in circles of radius $r$ in simulated trajectories of 1000 bouts in duration. The bold black line represents the average expected behavior over 3 groups. We find that $G_1$ is the best at long range exploration, while $G_2$ and $G_3$ are good at local exploitation.}
    \label{fig:S5_zebrafish_clus}
\end{figure*}

\begin{figure*}[ht!]
    \includegraphics{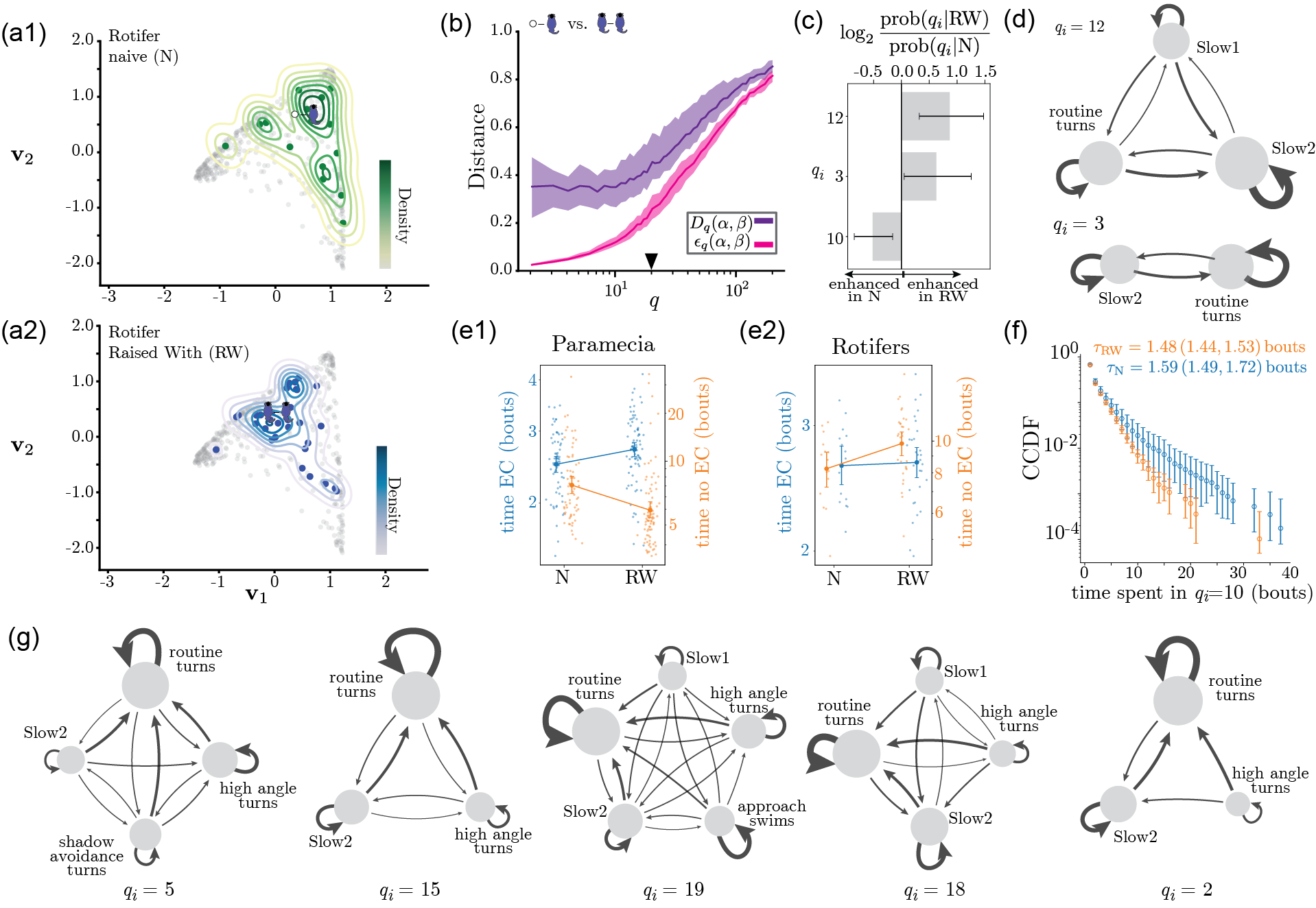}
    \caption{{\bf Details of the analysis of behavioral changes upon prior exposure in rotifers and paramecia.} (a) We compare naive (a1) vs experienced (a2) fish hunting rotifers. Prior exposures to rotifer shifts the distribution towards more exploratory phenotypes. (b) Distance $D_q$ and uncertainty $\varepsilon_q$ between naive and experienced fish hunting rotifers. While there is relevant information for a large number of of $q$, we compare conditions at $q=20$ as in Fig.\,\ref{fig:6}. (c) We report the log-fold change of the frequency of state usage at $q=20$ for statistically relevant differences with $\text{p-value}<0.05$ (two-sided permutation test) and $|\log_2\frac{\text{prob}(q_i|\text{RW})}{\text{prob}(q_i|\text{N})}|>0.5$. We find that raising fish with rotifers enhances the use of $q_i=12$ and $q_i=3$, while naive fish use more the $q_i=10$, which was enhanced in the Paramecia RW condition. (d) We plot the transition matrix among bout types for $q_i=12$ and $q_i=3$, which corresponds to sequences of Slow 1, Slow2 and Routine Turns, characteristic of exploratory behavior. (e) Bout sequence duration with (blue) or without (orange) eyes converged in the conditions in which fish is either hunting paramecia (e1) or rotifers (e2). Small dots represent the geometric mean bout sequence durations of each fish; larger markers and error bars show the average and its 95\% bootstrap confidence interval across fish. We find that fish raised with paramecia reduce the time spent with the eyes not converged (two-sided permutation test $\text{p-value}<0.001$) and increase the time spent with eyes converged (two-sided permutation test $\text{p-value}<0.01$). In contrast, when fish are raised with rotifers, we find that the time spent with eyes converged is unchanged (two-sided permutation test $\text{p-value}=0.8$), but the time spent with the eyes not converged increases upon raising the fish with prey (two-sided permutation test $\text{p-value}=0.028$). (f) Complementary Cumulative Distribution Function (CCDF) of the sequence length in state $q_i=10$ for the conditions in which the fish is hunting rotifers. Unlike fish hunting paramecia, Fig.\,\ref{fig:6}(e), we find that the sequence length in state $q_i=10$ is reduced after raising the fish with rotifers: the geometric mean duration was $\tau_\text{N} = 1.59\,(1.49,1.72)\,\text{bouts}$ for naive fish and $\tau_\text{RW} = 1.48\,(1.44,1.53)\,\text{bouts}$ for fish raised with paramecia, where values in parentheses indicate the 95\% confidence interval obtained by bootstrapping across fish. The difference between conditions was significant in a two-sided permutation test $\text{p-value}=0.036$. The CCDF was obtained using all the available bout sequences, and error bars represent 95\% confidence intervals bootstrapped across fish. (g) We plot the other transition matrices enhanced in the Paramecia Naive condition, Fig.\,\ref{fig:6}(c). They mostly constitute more exploratory bout sequences, with Slow1, Slow2, and variable turns. We also find $q_i=19$ with approach swims, which could represent a form of prey capture sequence used less after raising the fish with paramecia.}
    \label{fig:S6_zebrafish_rot}
\end{figure*}

\begin{figure*}[ht!]
    \includegraphics{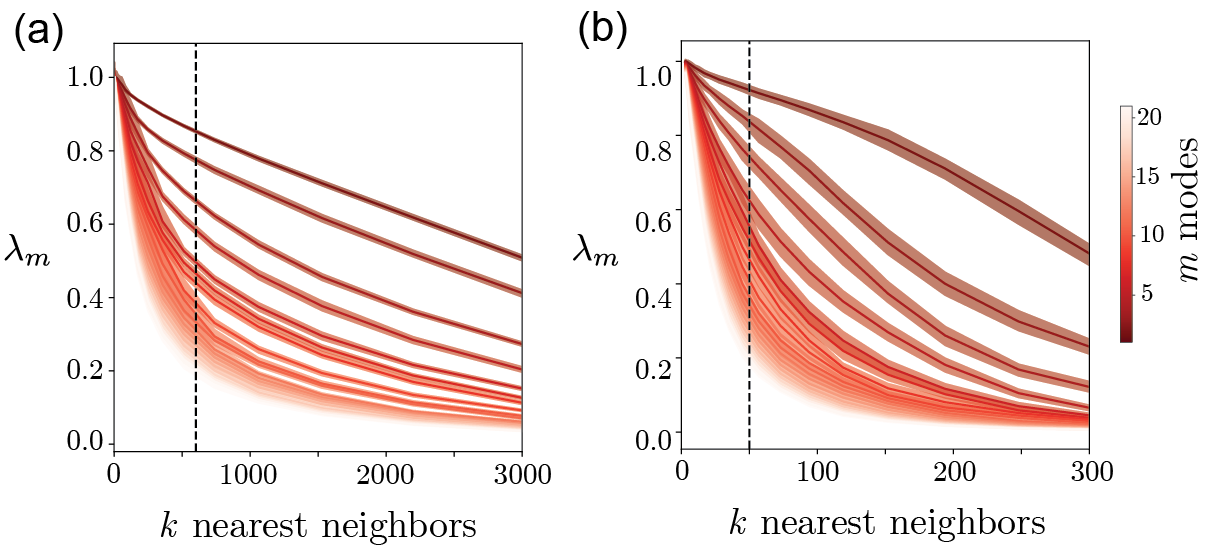}
    \caption{{\bf Selecting number of nearest neighbors $k$ for the self-tuning diffusion map for bacterial motility and larval zebrafish examples.} We calculate the eigenspectrum of the diffusion map for different values of $k$ for both systems. (a) Eigenvalues $\lambda_m$ for the bacterial locomotion phenotypic space showcasing the first $m$ eigenvalues of the system for multiple $k$. We select $k=600$ neighbors, at a point where the dominant modes separate more cleanly from the bulk. (b) Eigenvalues $\lambda_m$ for the larval zebrafish  phenotypic space showcasing the first $m$ eigenvalues of the system for multiple $k$. We select $k=50$, for which the eigenvalue separation is clear.}
    \label{fig:S7_dmap_k}
\end{figure*}

\clearpage




\section*{Diffusion maps capture the intrinsic geometry of phenotypic variation}

We consider that our time series observations, indexed by $\alpha$, result from a partial observation of an underlying Markovian stochastic process parameterized by $\theta^\alpha \in \mathbb{R}^N$ encoding underlying biological mechanisms. Changing in the underlying parameters map non-trivially to the emergent dynamics, and we use diffusion maps to probe the intrinsic geometry of the space of dynamics. Here, leveraging the connection between diffusion map eigenvectors and the Laplace-Beltrami operator \cite{coifman2006diffusion}, we show that the gradient of the diffusion map eigenvectors aligns locally with the dominant modes of the underlying metric tensor mapping variation in parameters to differences in dynamics.

We assume that in sufficiently local regimes the distance between observations $D$ admits a quadratic expansion,
\begin{equation}
    D(\theta,\theta+\delta)^2=\delta^\top g_D(\theta)\,\delta+o(\|\delta\|^2),\label{eq:D-local}
\end{equation}
This, in turn, induces a Riemannian metric tensor $g_D(\theta)$ on parameter space whose eigendirections define, locally in parameter space, which directions most impact the emergent dynamics. In addition, $g_D(\theta)$ defines a Laplace-Beltrami operator whose eigenfunctions capture the global intrinsic organization of the geometry induced by the distance between phenotypes, $D$. According to the Rayleigh-Ritz
variational characterization of Laplace-Beltrami eigenfunctions
\cite{chavel1984eigenvalues,petersen2006riemannian}, the non-trivial eigenfunctions of the Laplace-Beltrami operator, $f$, minimize the Dirichlet energy,
\begin{equation}
    \mathcal{E}[f]=\int(\nabla f)^\top g_D^{-1}(\nabla f)\,\mathrm{dvol}_g,\label{eq:dirichlet}
\end{equation}
subject to orthogonality constraints to lower modes. Consequently, low-frequency eigenfunctions vary preferentially along directions in parameter space where the induced geometry penalizes changes the least. This becomes clear when we project the gradient of the Laplace-Beltrami eigenfunctions locally in terms of the eigenspace of the underlying metric tensor $g_D(\theta)$. Let $g_D(\theta)e_a(\theta)=\Lambda_a(\theta)e_a(\theta),\,\Lambda_a\ge\Lambda_{a+1}>\ldots >0,$ be the local eigensystem of the metric tensor $g_D(\theta)$. Writing, $\nabla f=\sum_a \gamma_a e_a$, the energy density becomes
\begin{equation}
    (\nabla f)^\top g^{-1}(\nabla f)=\sum_a \frac{\gamma_a^2}{\Lambda_a}. \label{eq:Dirichlet_eigvals}
\end{equation}

Thus, in regions of parameter space where the metric $g_D(\theta)$ is strongly anisotropic, its dominant eigenvector $e_1(\theta)$ identifies the parameter direction along which the distance is locally most sensitive. Because the Dirichlet energy involves $g_D^{-1}$, gradients of low-frequency Laplace-Beltrami eigenfunctions can vary most cheaply along this same direction. We therefore expect the gradients of the leading diffusion-map eigenvectors to align locally with $e_1$.

\begin{figure*}[ht!]    
\includegraphics{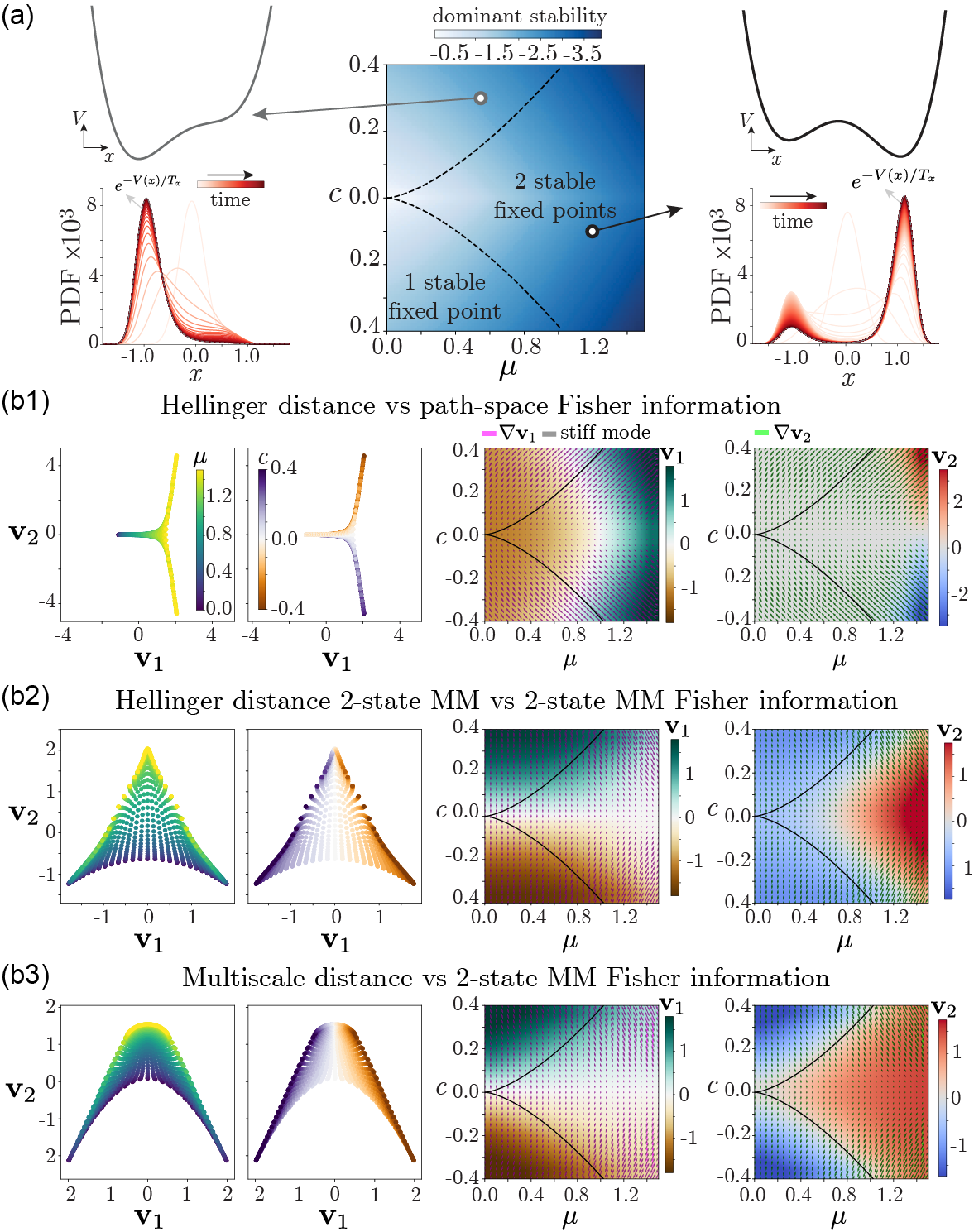}
    \caption{{\bf Relationship between diffusion map eigenvectors and the stiff directions of the induced metric mapping parameters to emergent dynamics.} 
    (a) - We consider overdamped Langevin dynamics in a quartic potential. In the center, we plot a phase diagram as a function of $\theta=(\mu,c)$, separating the regime with 2 stable fixed points from the regime with 1 stable fixed point at boundary $(c/2)^2=(\mu/3)^3$ (black dashed line). We color code the phase diagram by the curvature of the most stable fixed point. We also show examples of the potential landscape in the monostable (top) and bistable (bottom) regimes, as well as an example simulation of the density dynamics. (b) - We use diffusion maps to reconstruct the low-wavelength modes of the underlying Laplace-Beltrami operator, and showcase how the gradient of the diffusion map eigenvectors aligns with the stiff directions of the underlying metric. In each panel, we plot on the left the inferred diffusion map embedding, and on the right the parameter space color coded by each diffusion map eigenvectors, as well as the gradient of each eigenvector ($\nabla\mathbf{v}_1$ in pink and $\nabla\mathbf{v}_2$ in green) and the stiff mode of the induced metric tensor (grey). (b1) - Hellinger distance between paths, estimated using Eq.\,\ref{eq:hellinger-rate-distance}, compared against the path-space Fisher information matrix. We find that $\mathbf{v}_1$ is mostly associated with changes in $\mu$, and $\mathbf{v}_2$ is mostly associated with changes in $c$ (left).  The direction of change of the diffusion map eigenvectors matches the local stiff directions of the path-space Fisher information matrix, with an alignment score of $A_1=0.971\,(0.967,0.976)$ for $\mathbf{v}_1$ and $A_2=0.991\,(0.990,0.992)$ for $\mathbf{v}_2$. Notice that the stiff mode of the path-space Fisher information aligns with the change in stability of the most stable fixed point (shown in a). (b2) - Same as (b1), but now comparing the Hellinger distance between 2-state Markov models with the corresponding Fisher information matrix. We coarse-grain the dynamics into two states along $x=0$. We find that $\mathbf{v}_1$ mostly varies with $c$ whereas $\mathbf{v}_2$ mostly varies with $\mu$ (left). The gradient of the diffusion map agrees with the stiff mode of the underlying Fisher information matrix ($A_1=0.9992\,(0.9991,0.9993)$, $A_2=0.95\,(0.94,0.96)$). The dominant modes are orthogonal to the bifurcation boundary, separating monostable from bistable solutions. (b3) - Same as (b2), but now using the multiscale total variation distance introduced in the main text. The resulting embedding recovers the asymmetry of the potential landscape along $\mathbf{v}_1$, and the transition between bistable and monostable regimes along $\mathbf{v}_2$. The gradient of the diffusion map eigenvectors still agrees with the stiff modes of the Fisher information matrix computed from the Hellinger distance between 2-state Markov models ($A_1=0.979\,(0.976,0.982)$, $A_2=0.938\,(0.930,0.945)$).
    }
    \label{fig:SM1_cusp}
\end{figure*}

We test this prediction in a cusp catastrophe model, Fig.\,\ref{fig:SM1_cusp}, and illustrate how the result depends on the choice of distance $D$: different notions of distance between the stochastic dynamics induce different metric tensors and therefore identify different directions in parameter space as dynamically important. 

We study a stochastic cusp catastrophe model,
\begin{equation}
    \mathrm{d}X_t=-\partial_x V(X_t;\mu,c)\,\mathrm{d}t+\sqrt{2T_x}\,\mathrm{d}W_t, \text{where } V(x;\mu,c)=\frac{x^4}{4}-\mu\frac{x^2}{2}+cx,\label{eq:cusp_SDE}
\end{equation}

which exhibits either one or two fixed points depending on $\mu$ and $c$, and is either right or left biased depending on the sign of $c$,Fig.\,\ref{fig:SM1_cusp}(a). We first consider a path-space distance, for which both the local Hellinger geometry and the associated path-space Fisher information rate can be determined analytically. Let $\mathbb{P}_\theta$ denote the path measure generated by the Langevin dynamics at parameter value $\theta$. Since two parameter values $\theta,\theta'$ differ only in drift, while sharing the same noise strength, Girsanov theory gives the local path-space Fisher information rate,
\begin{equation}
    \frac{1}{T}
    \mathrm{KL}\left(\mathbb{P}_{\theta}\,\|\, \mathbb{P}_{\theta+\delta}\right)
    =
    \frac12\delta^\top I_{\mathrm{path}}(\theta)\delta+o(\|\delta\|^2),
\end{equation}
where
\begin{equation}
    I_{\mathrm{path}}(\theta)_{ij}=\frac{1}{2T_x}\mathbb{E}_{\pi_{\theta}}
    \left[
        \partial_{\theta_i}\partial_x V(X;\theta)\,
        \partial_{\theta_j}\partial_x V(X;\theta)
    \right].n\label{eq:Ipath_potential}
\end{equation}
For the quartic potential, we have,
\begin{equation}
    I_{\mathrm{path}}(\mu,c)=\frac{1}{2T_x}
    \begin{pmatrix}
    \mathbb{E}_{\pi_{\mu,c}}[X^2] &-\mathbb{E}_{\pi_{\mu,c}}[X]\\
    -\mathbb{E}_{\pi_{\mu,c}}[X]&1
    \end{pmatrix},\label{eq:Ipath_cusp}
\end{equation}
where $\mathbb{E}_{\pi_{\mu,c}}$ denotes expectation under the stationary density $\pi_{\mu,c}\propto e^{-V(x;\mu,c)/T_x}$. Although the Fisher tensor is naturally derived from the local expansion of the KL divergence, the KL divergence is asymmetric and is therefore not ideal as a distance for constructing diffusion-map kernels. We instead use the corresponding local Hellinger-rate distance, which is symmetric and induces the same local Fisher geometry up to a constant factor. For two parameter values $\theta_a=(\mu_a,c_a)$ and $\theta_b=(\mu_b,c_b)$, let $\theta_m=(\theta_a+\theta_b)/2$. We define the symmetric midpoint approximation
\begin{equation}
    D_H^2(\theta_a,\theta_b)=\frac{1}{16T_x}\int\left[(\mu_b-\mu_a)x-(c_b-c_a)\right]^2\pi_{\theta_m}(x)\,dx.
    \label{eq:hellinger-rate-distance}
\end{equation}
This expression is the local quadratic approximation to the squared Hellinger distance per unit time between path measures. Indeed, for infinitesimal perturbations,
\begin{equation}
    D_H^2(\theta,\theta+\delta)=\frac{1}{8}\delta^\top I_{\mathrm{path}}(\theta)\delta+o(\|\delta\|^2).\label{eq:DH-Ipath}
\end{equation}
Thus, the path-space Hellinger-rate distance is Fisher-local: it has the same eigendirections as $I_{\mathrm{path}}(\theta)$. Therefore, following the argument of Eqs.\,\ref{eq:D-local},\ref{eq:dirichlet},\ref{eq:Dirichlet_eigvals}, we expect the diffusion map eigenvectors computed from Hellinger-rate distances to vary along the stiff directions of the path-space Fisher information matrix. Fig.\,~\ref{fig:SM1_cusp}(b1) tests the resulting prediction. At each point in parameter space, we compare the gradient of a diffusion-map eigenvector with the dominant eigenvector of the local Fisher metric, and quantify their alignment by
\begin{equation}
    A_k = \sum_i \omega_i \frac{ \left| \nabla_\theta \mathbf{v}_k(\theta_i) \cdot e_1(\theta_i) \right|}{\|\nabla_\theta \mathbf{v}_k(\theta_i)\|},
    \label{eq:alignment-score}
\end{equation}
where $\omega_i\propto \frac{\Lambda_1(\theta_i)}{\Lambda_2(\theta_i)}$ emphasizes regions in which the metric is strongly anisotropic. The gradients of the leading diffusion coordinates show strong local alignment with the dominant Fisher direction, as predicted by the geometric argument above. The stiff mode of the path-space Fisher information rate is aligned parallel to the bifurcation boundary, highlighting how the Hellinger distance between paths is most sensitive to infinitesimal changes in parameters integrated over the entire domain: the stiff mode in this case seems to vary with the stability of the most stable minimum of the potential, Fig.\,\ref{fig:SM1_cusp}(a).

We next showcase how different distances highlight different properties of the dynamics, while the geometric argument holds. For example, we consider the Hellinger distance among 2-state Markov models, which compresses the finescale dynamics into discrete labels depending on the sign of $x$. In this case, we find that the stiff Fisher mode is orthogonal to the bifurcation boundary, highlighting how at the level of a 2-state description of the system this is the direction that most changes the transition dynamics at the longest timescales, Fig.\,\ref{fig:SM1_cusp}(b2). Accordingly, the diffusion map eigenmodes vary with $\mu$ and $c$ but along directions that directly capture the transition between bistable and monostable solutions, and their gradients align with the Fisher information eigenvectors. 

For completeness, we also assess the embedding obtained using our multiscale dissimilarity, Eq.\,\ref{eq:D_multiscale}. Despite being agnostic about the correct coarse-graining scale to compare the dynamics, our approach recovers the alignment of the diffusion map eigenvectors with the bifurcation structure and the asymmetry of the potential landscape. In addition, we find a very good agreement between the gradient of the diffusion map eigenvectors and the Fisher information metric induced by the Hellinger distance among 2-state Markov chains, indicating that our multiscale metric preserves the most relevant long-lived structure in the dynamics in an unbiased way.

\section*{Comparing the spectra of transfer operators}

Beyond the path-space Fisher information rate, another common approach for comparing dynamical systems is to compare the spectra of their transfer operators \cite{ostrow2023beyond,huang2025inputdsa,kamiya2024koopman,zhang2025koopstd}. In particular, the optimal transport distances between their singular-value spectra provides a compact measure of changes in dynamical structure, including changes in the number and stability of fixed points. Our framework provides a convenient way to compute such distances because the construction of our coarse-grained states already relies on Ulam-Galerkin approximations of Perron-Frobenius operators (Methods). We can therefore directly compare the singular value spectra of these operators using a Wasserstein distance.

The Lorenz system \cite{lorenz1963deterministic} illustrates the strength of this spectral comparison, Fig.\,\ref{fig:SM2_wasserstein}(a,b,c). As parameters are varied, the Wasserstein distance between transfer operator spectra distinguishes qualitatively different dynamical regimes, including fixed points, limit cycles, and chaotic dynamics, allowing these phases to be recovered in the corresponding diffusion map embedding.

The stochastic cusp model, however, exposes an important limitation of this representation, Fig.\,\ref{fig:SM2_wasserstein}(d). Applied to Eq.\,\ref{eq:cusp_SDE}, the resulting embedding is primarily sensitive to changes in $\mu$, but does not distinguish $\theta=(\mu,c)$ from $\theta=(\mu,-c)$. Changing the sign of $c$ reflects the potential about $x=0$. The two systems therefore have the same dynamical organization, including the same number and stability of fixed points, while exchanging the identities of the left- and right-biased states. Their transfer-operator spectra consequently provide no information about which of these states is being modulated. For some applications this invariance is desirable: it identifies systems with the same underlying dynamical structure irrespective of the labeling or location of their states. For dynamical phenotypes, however, the identity of the affected state can itself be biologically meaningful. In comparison, our multiscale dissimilarity retains sensitivity to changes in metastable structure while also keeping track of which dynamical states are changing.

\section*{Details of numerical analysis}\label{sec:Sim_details}

\subsection*{Stochastic cusp catastrophe}

Instead of making simulations, we here directly approximate the infinitesimal generator of the Fokker-Planck dynamics using the Chang-Cooper method \cite{chang1970practical}, discretizing the domain $x\in [-\sqrt{\mu_\text{max}}-4\sqrt{T_x}, \sqrt{\mu_\text{max}}+4\sqrt{T_x}]$ uniformly into 1000 points, where $\mu_\text{max}$ is the largest sampled value of $\mu$. The value of $\mu$ was sampled between 0 and 1.4, with a uniform spacing of $\Delta\mu=0.05$, whereas $c$ values were sampled between -0.4 and 0.4, with a uniform spacing of $\Delta c = 0.025$.

\begin{table}[ht!]
    \centering
    \begin{tabular}{c|c}
          dynamical regime & $r$ \\
         \hline
        stable fixed point & 20 \\
        periodic $[1,1]$  & 220\,;\,250 \\
        periodic $[1,2,2]$ & 100.5 \\
        periodic $[1,2,2]^2$ & 99.7 \\
        periodic $[1,2,2]^4$ & 99.6 \\
        periodic $[1,2,2]^8$ & 99.537 \\
        periodic $[1,1,2,2]$ &  152\,,\,165 \\
        periodic $[1,1,2,2,1,2$ & 126.5 \\
        intermittent chaos & 100.93 \\
        chaos & 28\,;\,45\,;\,60\,;\,75\,;\,99.4\,;\,115\,;\,140\,;\,195 \\
    \end{tabular}
    \caption{Sampled dynamical regimes of the Lorenz system across $r$.}
    \label{tab:placeholder}
\end{table}

\subsection*{Lorenz system}

\begin{figure*}[ht!]
    \includegraphics{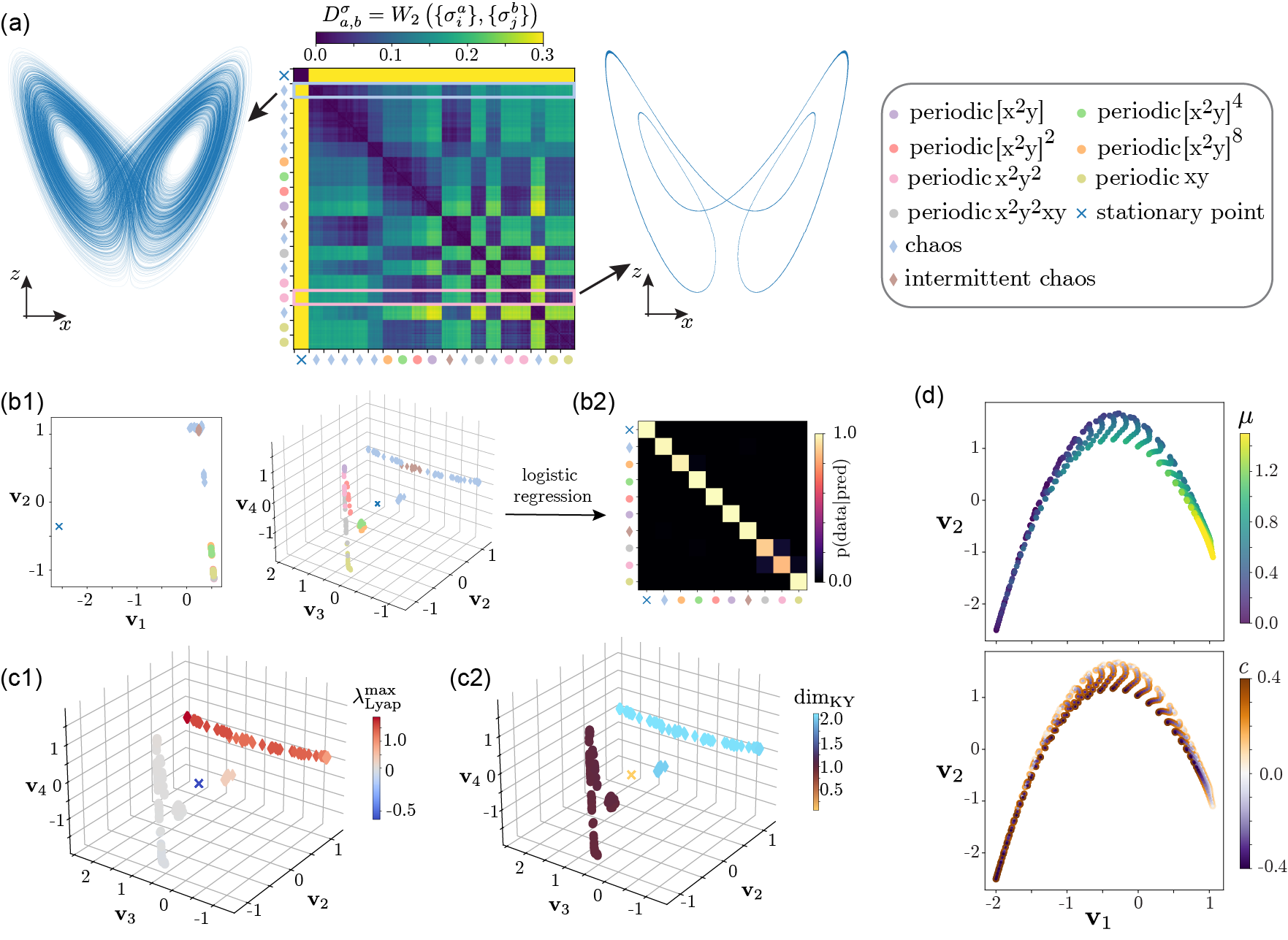}
    \caption{{\bf Comparing dynamics using the optimal transport distance between the singular values of the inferred transfer operators.} (a) Our approach provides an efficient approximation of Perron-Frobenius operators from time series data. We here showcase how the optimal transport distance among their singular values enables the recovery of different dynamical regimes in the Lorenz system \cite{lorenz1963deterministic}, Eq.\,\ref{eq:Lorenz}. We set $b = 8/3$, $\sigma=10$ and vary $r$ to sample different dynamical regimes. We simulate 10 time series from different initial conditions for each value of $r$, and show example simulations (left: chaos $r=28$, right: periodic solution $r=165$). We estimated the Wasserstein distance between the singular values of an inferred Perron-Frobenius operators, $D_{a,b}^\sigma = W_2(\{\sigma_i^a\},\{\sigma_j^b\})$ (center). The tick marks represent different values of $r$. (b) Diffusion map embedding of the Wasserstein distance matrix. (b1) The first dimension $\mathbf{v}_1$ isolates out the stable fixed point solution from everything else (left); $\mathbf{v}_2$ captures the difference between limit cycle and chaotic solutions, $\mathbf{v}_3$ differences among chaotic solutions, and $\mathbf{v}_4$ differences among periodic solutions. (b2) We use a regularized logistic regression to classify the points in the diffusion map embedding into their respective dynamical regimes, resulting in classification accuracy of $0.97\pm 0.04$ (see Section \ref{sec:Sim_details}). We plot the average confusion matrix obtained from repeated stratified 5-fold cross-validation (100 repetitions). (c) We examine how the different diffusion map dimensions map onto properties of the Lyapunov spectrum. (c1) We color code the diffusion map embedding by the largest Lyapunov exponents, showcasing how $\mathbf{v}_2$ distinguishes chaotic from non-chaotic dynamics, while $\mathbf{v}_3$ subsequently captures the degree of chaos. (c2) We color code the diffusion map embedding by the Kaplan-Yorke dimension \cite{frederickson1983liapunov}, showcasing the difference between fixed point solutions (which lie along negative $\mathbf{v}_1$, visible in panel b1), limit cycle solutions and fractal chaotic solutions. 
    (d) We reproduce the results on the cusp catastrophe example Fig.\,\ref{fig:SM1_cusp}, but using a Wasserstein distance between singular values of $\mathcal{L}_\theta$. We approximate $\mathcal{L}_\theta$ using the Chang-Cooper method \cite{chang1970practical} for each $\theta=(\mu,c)$, and compute the singular values of $\mathcal{P}_\tau = e^{\mathcal{L}\tau}$ with $\tau=0.1\,\text{s}$. The diffusion map embedding constructed with a self-tuning kernel (with 4 nearest neighbors), shows that $\mathbf{v}_1$ mostly aligns with variation in $\mu$, while $\mathbf{v}_2$ can capture the overall asymmetry of the potential, but cannot distinguish $\theta=(\mu,c)$ from $\theta=(\mu,-c)$ because these are topologically conjugate.
    }
    \label{fig:SM2_wasserstein}
\end{figure*}

We simulate time series from the equations of motion,

\begin{align}
\begin{cases}
    \dot{x} & = \sigma(x-y) \\
    \dot{y} & = (r-z)x-y \\ 
    \dot{z} & = xy-\beta z, 
\end{cases}\label{eq:Lorenz}
\end{align}
with $\sigma=10$, $\beta=8/3$ and varying $r$ sampled to capture different dynamical regimes \cite{sparrow2012lorenz} as in \cite{kamiya2024koopman} (see Table I). We sampled from each value of $r$ 10 times, using the {\tt scipy.odeint} package \cite{2020SciPy-NMeth} to simulate for $T_\text{expt}=1000\,\text{s}$ with a sampling time of $\delta t = 0.01\,\text{s}$, discarding the first $1000\,\text{s}$ as transient dynamics. We then approximated the Perron-Frobenius operator as in \cite{costa2023maximally} by partitioning the state space into $N=1000$ partitions and using a transition time of $\tau=0.1\,\text{s}$, obtaining a $N\times N$ transition operator $P_{ij}$ for each simulated timeseries. Finally, we estimated the Wasserstein distance between the singular values $\{\sigma_i\}$ of each $P_{ij}$, Fig.\,\ref{fig:SM2_wasserstein}(a), and used self-tuning diffusion maps with a bandwidth inferred from 30 nearest neighbors to reconstruct the phenotypic space in Figs.\,\ref{fig:SM2_wasserstein}(b,c). We trained an L1-regularized multinomial logistic regression classifier to predict the dynamical regimes from the diffusion map coordinates (keeping 20 dimensions). The classifier performance was assessed using repeated stratified 5-fold cross-validation with 100 repeats, and the confusion matrix shown in Fig.\,\ref{fig:SM2_wasserstein}(b2) is averaged across all 500 splits. We used the Lyapunov spectrum to access how the diffusion map dimensions resolve different properties of the dynamics: unpredictability (as measured through the largest Lyapunov exponent), and dimensionality (as measured through the Kaplan-Yorke dimension \cite{frederickson1983liapunov}). Lyapunov spectra were estimated using a continuous Gram–Schmidt method, integrating the analytical Jacobian over $1000\,\text{s}$. For each parameter $r$, we repeated the calculation 5 times and averaged the resulting spectra over realizations.

\end{document}